\documentclass{IEEEtran}
\usepackage{cite}
\usepackage{amsmath,amssymb,amsfonts}
\usepackage{algorithmic}
\usepackage{graphicx}
\usepackage{textcomp}

\usepackage{subfigure}

\usepackage{comment}

\usepackage{tablefootnote}

\let\emptyset\varnothing
\usepackage{booktabs}
\usepackage{makecell}

\usepackage[linesnumbered]{algorithm2e}
\SetKwComment{Comment}{$\triangleright$\ }{}
\RestyleAlgo{ruled}

\usepackage{mathtools}
\DeclarePairedDelimiter{\ceil}{\lceil}{\rceil}

\allowdisplaybreaks

\usepackage{tikz}

\begin{document}


\title{Hybrid Quantum-Classical Multi-cut \\ Benders Approach with a Power System Application}

\author{Nikolaos G. Paterakis,~\IEEEmembership{Senior Member,~IEEE}
\thanks{N.G. Paterakis is with the Department of Electrical Engineering, Eindhoven University of Technology, Eindhoven, 5600MB, The Netherlands (email: n.paterakis@tue.nl).
}
}
\renewcommand\footnoterule{\hspace{-1em}\rule[0.45em]{\columnwidth}{0.45pt}}



\maketitle

\begin{abstract}
Leveraging the current generation of quantum devices to solve optimization problems of practical interest necessitates the development of hybrid quantum-classical (HQC) solution approaches. In this paper, a multi-cut Benders \mbox{decomposition (BD)} approach that exploits multiple feasible solutions of the master problem (MP) to generate multiple valid cuts is adapted, so as to be used as an HQC solver for general mixed-integer linear programming (MILP) problems. The use of different cut selection criteria and strategies to manage the size of the MP by eliciting a subset of cuts to be added in each iteration of the BD scheme using quantum computing is discussed. The HQC optimization algorithm is applied to the Unit Commitment (UC) problem. UC is a prototypical use case of optimization applied to electrical power systems, a critical sector that may benefit from advances in quantum computing. The validity and computational viability of the proposed approach are demonstrated using the D-Wave Advantage 4.1 quantum annealer.
\end{abstract}

\begin{IEEEkeywords}
Benders decomposition, power systems, quantum computing, unit commitment
\end{IEEEkeywords}

\section{Introduction}

\IEEEPARstart{Q}{uantum} computing (QC) is an emerging technology that has the potential to help addressing hard computational problems. Although significant progress has been made in the recent years, universal error-corrected quantum computers have not been realized yet. Nonetheless, currently available quantum hardware, often called Noisy Intermediate-Scale Quantum (NISQ) \cite{Preskill_2018}, allows applying quantum algorithms in order to identify problems and application areas in which QC can offer an advantage with respect to classical computing. Among the many areas where a quantum speedup is expected, solving combinatorial optimization problems is one of the most prominent ones \cite{Lucas_2014}. 

In the context of using QC for optimization, a commonly-studied class of combinatorial problems are Quadratic Unconstrained Binary Optimization (QUBO) problems \cite{Kochenberger_2014}. A QUBO problem is represented by $\min_{\mathbf{x}} \mathbf{x}^{T} \mathbf{Q} \mathbf{x}$ where \mbox{$\mathbf{x} \in \left\{0,1 \right\}^n$} and $\mathbf{Q} \in \mathbb{R}^{n \times n}$. QUBO problems can be transformed to Ising models ($\mathbf{x} \in \left\{-1,1\right\}^n$) and vice versa. The limitations of NISQ hardware have triggered the development of hybrid quantum-classical (HQC) algorithms that exploit both classical and quantum resources. Popular techniques include variational approaches such as the Variational Quantum Eigensolver (VQE) \cite{Peruzzo_2013} and the Quantum Approximate Optimization Algorithm (QAOA) \cite{Farhi_2014}, as well as techniques based on Grover's algorithm \cite{Gilliam_2021}. However, gate-based QC, which the aforementioned algorithms are designed to exploit, are still at an early development stage and, therefore, their application for solving practical-scale problems is limited \cite{Nannicini_2019}. On the contrary, special-purpose quantum computers, namely quantum annealers (QA), currently surpass gate-based computers both in terms of number of qubits and qubit connectivity and can be used to obtain approximate solutions of larger QUBO instances \cite{McGeoch_2020}. 

\subsection{Decomposition-based HQC Algorithms}

Although several problems can be modeled as QUBO (e.g., \cite{Grant_2021,Harwood_2021,Stollenwerk_2019}), most optimization problems of practical interest contain both discrete and continuous variables. For instance, mixed-integer linear programming (MILP) problems are commonly used in diverse application areas, such as \mbox{logistics \cite{Anghinolfi_2016_example_logistics},} the coordination of unmanned aerial \mbox{vehicles \cite{Yang_2019_example_UAV}} and power systems \cite{Chen_2016_example_powersystemMILP}. \looseness=-1 

In order to leverage a quantum processing unit (QPU) for solving mixed-integer problems of practical interest, decomposition-based HQC algorithms must be developed. Such HQC algorithms aim to decompose the original optimization problem into a part that can be assigned to the QPU, i.e., a QUBO problem or a problem that can be cast as a QUBO, and a part that can be solved efficiently using classical algorithms (e.g., a convex optimization problem). Recently, the development of such approaches has gained attention and various general-purpose and problem-specific techniques have been proposed. \looseness=-1 

In \cite{Gambella_2020} a decomposition approach for mixed binary-continuous optimization problems with complicating constraints based on a multi-block version of the alternating direction method of multipliers (ADMM) was presented. The proposed method splits the original problem into a QUBO problem and constrained convex subproblems (SP). The QUBO problem was solved using the variational quantum eigensolver (VQE) and the quantum approximate optimization algorithm (QAOA), while the constrained convex SPs were solved with a classical solver. Despite being a general method of wide applicability, the convergence of the HQC is not guaranteed. \looseness=-1 

In \cite{Chang_2020} the Benders decomposition (BD) scheme was used as the basis for a general-purpose HQC MILP solver. BD splits the problem into a master problem (MP) that contains both binary and continuous variables and a linear programming (LP) SP. A primer on Benders decomposition is presented in \mbox{Section \ref{Benders_decomposition}.} Although BD is proven to converge to the global optimum of the original MILP problem, a major drawback of its direct application as an HQC algorithm is that it requires the discretization of the continuous variable that appears in the constraints of the MP (proxy for the SP value) at the expense of using ancillary qubits. 


Apart from the aforementioned general-purpose HQC algorithms, that are based on popular decomposition techniques, similar ideas have been proposed for specific optimization problems (e.g., \cite{Ajagekar_2020, Braine_2021}).

\subsection{QC Applications in the Power \& Energy Sector} \label{intro_2}

Although a general-purpose decomposition-based HQC optimization approach is proposed in this paper, this study is motivated by a critical application area, namely power and energy systems. Computing has always been of paramount importance for the design and operation of power \mbox{systems \cite{Zobaa_2018}.} However, the complexity induced by the modernization of the power sector is posing computational challenges that may not be met by classical resources \cite{Facio_2021}. The energy sector is currently undergoing a transition from fossil-based to zero-carbon energy sources, triggered by global decarbonization targets. The electrical power sector is leading the energy transition through three innovation trends, namely the electrification of end-use sectors, the decentralization of energy resources and extensive digitalization. It is foreseen that electricity will become the main energy carrier due to the significant potential of integrating renewable energy resources in power systems, while the linkage with other energy carriers (e.g., transportation and gas) will be strengthened, resulting in an even more complex \mbox{system \cite{Potrc_2021}.} 

In this context, QC and quantum informatics are expected to find significant applications in the energy sector. First, quantum cryptography can be leveraged in order to shield the cyber security of power systems, especially those with many distributed energy resources \cite{Tang_2020}. Also, QC has the potential to expand the current computational capabilities and facilitate the solution of complex analysis, optimization and data analytics problems in the energy domain \cite{Eskandarpour_2020_survey,Obafemi_2021,Giani_2021}, while maintaining a low energy footprint of \mbox{computation \cite{Elsayed_2019}.} Interestingly, although the prospect of benefits has been identified, technical studies that investigate the use of QC for addressing power and energy system computational problems are scarce. 

In \cite{Jones_2020} the solution of the phasor measurement unit (PMU) placement problem using the D-Wave Systems 2000Q QA was investigated. The PMU placement problem was formulated as the minimum dominating set problem. It was found that for some instances the QA outperformed the classical solver CPLEX. The optimization model presented in \cite{Jones_2020} is easily converted into a QUBO, however, it does not capture the complexity of the actual problem. Similarly, in \cite{Loh_2019} fairness during the design of a heat grid was evaluated by solving a QUBO problem on a QA and the quality of the solutions was benchmarked against a graph partitioning solver without observing significant differences. In \cite{Ajagekar_2019} simplified formulations of the facility location-allocation, \mbox{unit commitment (UC)} and heat exchanger network synthesis problems were solved using both the D-Wave Systems 2000Q QA and \mbox{IBM Q} gate-based quantum computers. It was reported that for some problem instances, contrary to the QC approach, the classical solver Gurobi failed to return an optimal solution within a given time limit. However, the problem formulations presented \mbox{in \cite{Ajagekar_2019}} could either be directly recast as QUBO problems or discretization of continuous variables was performed, at the expense of using a large number of ancillary qubits in order to represent the discretized variables. 

Quantum algorithms have also been applied to fundamental power system analysis. In \cite{Feng_2021} a quantum power flow algorithm that exploits the Harrow-Hassidim-Lloyd (HHL) algorithm for the solution of systems of linear equations was introduced. Although computational experiments are performed on a small-scale test system due to the limitations of NISQ technology, it was argued that an exponential quantum speedup may be attainable in the future. The HHL algorithm was also exploited in \cite{Eskandarpour_2020} as part of a workflow to assess power system security under contingencies of increasing severity. These two studies advocate that QC may find impactfull applications in risk and reliability assessment of power grids, that are currently considered intractable for complex large-scale systems. \looseness=-1 

Finally, applications of HQC algorithms that combine machine learning and quantum sampling have been applied to power system problems. For instance, in \cite{Ajagekar_2021_1} a methodology to detect and classify power system faults using conditional restricted Boltzman machines and quantum generative training was proposed. \looseness=-1 

\subsection{Contributions}

The convergence properties of BD are very appealing, however, previous research proposed a straightforward implemention of the method as the basis of an HQC optimization algorithm that does not efficiently exploit NISQ \mbox{resources \cite{Chang_2020}.} In this paper, an alternative approach to implementing BD is investigated. Instead of discretizing the continuous variable of the MP in order to directly assign it to a QPU, a multi-cut version of BD is developed. The QPU is assigned to handle a pure binary subroutine that selects the cuts that enter the MP in order to manage its size and accelerate BD, while both the MP and the SPs are solved on a classical computer. 

The contribution of this paper is threefold:
\begin{itemize}
    \item An HQC multi-cut BD scheme that is applicable to general MILP problems without decomposable SP structure is adapted such that both classical and quantum resources can be exploited.
    \item A cut selection procedure that is based on two different cut selection criteria (exclusion of infeasible MP solutions, MP variable coverage) and two different cut selection strategies (minimum set cover, maximum coverage) is proposed. The QUBO reformulation of the cut selection strategies is provided and implementation details are discussed. \looseness=-1 
    \item The proposed HQC optimization algorithm is applied to a prototypical power system optimization problem, namely the UC problem, and the computational viability of its solution using a commercially available QA is extensively discussed. \looseness=-1 
\end{itemize}

It is to be noted that although in this paper experiments are conducted using a QA, the quantum step can also be solved on a gate-based QPU using, for instance, VQE and QAOA.

The remainder of this paper is organized as follows: in \mbox{Section \ref{lab:background}} the necessary theoretical background is established, while in \mbox{Section \ref{lab:methodology}} the proposed HQC multi-cut BD algorithm and the proposed cut selection procedure are detailed. Then, in \mbox{Section \ref{lab:use_cases}} the UC problem formulation that is used for demonstration purposes is described. Numerical results are presented and discussed in \mbox{Section \ref{lab:numerical_experiments}.} Finally, conclusions are drawn in \mbox{Section \ref{lab:conclusion}}.

\section{Preliminaries} \label{lab:background}
\subsection{Quantum Annealing} \label{Quantum_annealing}

In this section, an overview of the process of solving an optimization problem using a QA is briefly reviewed. Further details can be found in various sources, including \cite{McGeoch_2020} \mbox{and \cite{Johnson_2011}.} QAs are based on the adiabatic quantum computing \mbox{paradigm \cite{Childs_2011}.} Quantum annealing evolves a quantum state by applying a time-dependent Hamiltonian described by \eqref{eq:annealing_1} over a time interval $[0,T_{A}]$, where $T_{A}$ is the annealing time:

\begin{equation}
    H(\tau) = A(\tau)H_{0} + B(\tau)H_{P}   \label{eq:annealing_1}
\end{equation}

The annealing path functions $A(\tau)$ and $B(\tau)$ are monotonic functions of time that satisfy the conditions $A(0)=1$, $A(T_{A})=0$ and $B(0)=0$, $B(T_{A})=1$ respectively. In other words, at the beggining of the annealing, the Hamiltonian is equivalent to $H_{0}$ and gradually transitions to $H_{P}$. The initial Hamiltonian $H_{0}$ for a system of $V$ qubits is described by \eqref{eq:annealing_2}, where $\sigma^{x}_{a}$ is the Pauli-$x$ operator applied to qubit $a$. The initial Hamiltonian sets the qubits into an equal superposition with respect to the computational basis $z$ (even representation of all the optimization problem solutions). 

\begin{equation}
    H_{0} = -\sum_{a \in V} \sigma^{x}_{a} \label{eq:annealing_2}
\end{equation}

The problem Hamiltonian $H_{P}$ which is dominant at time $T_{A}$ is described by \eqref{eq:annealing_3} and represents the unconstrained optimization problem of interest in terms of an Ising model, whose ground state is the optimal solution. The parameters $h$, $J$ are real numbers that depend on the problem to be solved (linear and quadratic biases) and $\sigma_{a}^{z}$ is the Pauli-$z$ operator applied to qubit $a$.

\begin{equation}
    H_{P} = \sum_{a \in V} h_a \sigma_{a}^{z} + \sum_{a \in V}\sum_{b \in V, a \neq b} J_{a,b} \  \sigma_{a}^{z} \otimes \sigma_{b}^{z} \label{eq:annealing_3}
\end{equation}

Ideally, according to the quantum adiabatic theorem, starting from the ground state of $H$, if the transition $A: 1 \rightarrow 0$ and $B: 0 \rightarrow 1$ is sufficiently slow, the system will remain in the ground state throughout the transition. This implies that $H(T_{A})$ should also be in the ground state, i.e., it represents the optimal solution of the \mbox{problem \cite{Farhi_2000}.} 

Since QAs are open systems, the final result may not be the optimal solution of the problem. For this reason, quantum annealing is applied multiple times (annealing-read cycle) in order to increase the probability of finding high quality solutions. In general, the process of solving an optimization problem using a QPU consists of three steps. First, the optimization problem must be expressed as an Ising model or, equivalently, a QUBO problem. It is possible that the logical problem formulation may impose interactions between qubits that are not directly compatible with the physical topology of the QPU. For this reason, the second step consists of finding a minor embedding of the logical problem graph such that it is compatible with the sparse native topology of the QPU, i.e., the physical connectivity between qubits \cite{Cai_2014}. Finally, the minor-embedded problem instance is submitted to the QPU together with a set of hyperparameters, QA is performed and a set of solutions is returned.

\subsection{Benders Decomposition}\label{Benders_decomposition}

In this paper, we are concerned with MILP problems of the form \eqref{eq:MILP_problem}: 
\begin{subequations}
\begin{alignat}{2}
&\!\min_{\mathbf{x},\mathbf{y}}        &\qquad& \mathbf{c}^{T}\mathbf{x}+\mathbf{d}^{T}\mathbf{y}\label{eq:original_objfun}\\
&\text{subject to} &                          & \mathbf{A}\mathbf{x} + \mathbf{B}\mathbf{y}  \geq \mathbf{b}\label{eq:original_complicating}\\
&                  &                          & \mathbf{x}\in \mathbb{R}^{n}_{+}, \mathbf{y}\in \mathbb{Y}^{m} \label{eq:original_domain}
\end{alignat}\label{eq:MILP_problem}
\end{subequations}

\vspace{-10pt}\noindent where $\mathbf{c} \in \mathbb{R}^{n}$, $\mathbf{d} \in \mathbb{R}^{m}$, $\mathbf{b} \in \mathbb{R}^{q}$ are vectors and $\mathbf{A} \in \mathbb{R}^{q \times n}$, $\mathbf{B} \in \mathbb{R}^{q \times m}$ are matrices of coefficients, respectively. $\mathbb{Y}$ is a set of constraints involving only the decision variables $\mathbf{y}$.

BD is a popular technique that is applied to problems with complicating decision variables such as \eqref{eq:MILP_problem} \cite{Conejo_2006}. Decision variables may be considered complicating if they are involved in most of the problem constraints or render the optimization problem non-convex. If complicating variables are fixed, then the optimization problem is substantially simplified, either because it can be decomposed in SPs that can be solved independently, or because matrix $\mathbf{A}$ attains a structure which permits a straightforward solution. The central idea in BD is to decompose the original problem into an MP which contains all the integer variables and a SP in which the complicating variables $\mathbf{y}$ are fixed to tentative values (LP problem). The MP and the SP are iteratively solved. Tentative solutions $\hat{\mathbf{y}}$ are provided by solving the MP, whereas the solution of the SP yields the so-called Benders cuts, i.e., constraints that are progressively added to the MP to restrict its solution space. 

Given a tentative solution of the MP ($\mathbf{y}=\hat{\mathbf{y}}$) the SP is given by \eqref{eq:BD_original_SP}: \vspace{-15pt}

\begin{subequations}
\begin{alignat}{2}
&\!\min_{\mathbf{x}}        &\qquad& \mathbf{c}^{T}\mathbf{x}\label{eq:sp_objfun}\\
&\text{subject to} &               & \mathbf{A}\mathbf{x} \geq \mathbf{b}-\mathbf{B}\hat{\mathbf{y}} \label{eq:sp_constraint}\\
&                  &               & \mathbf{x}\in \mathbb{R}^{n}_{+} \label{eq:sp_domain}
\end{alignat}\label{eq:BD_original_SP}
\end{subequations}

\vspace{-15pt}In practice, the dual of the SP (DSP) given by \eqref{eq:BD_original_DSP} is solved: \vspace{-15pt}

\begin{subequations}
\begin{alignat}{2}
&\!\max_{\mathbf{v}}        &\qquad&  V_{DSP}=(\mathbf{b}-\mathbf{B}\hat{\mathbf{y}})^T \mathbf{v} \label{eq:dsp_objfun}\\
&\text{subject to} &               & \mathbf{A}^T\mathbf{v} \leq \mathbf{c} \label{eq:dsp_constraint}\\
&                  &               & \mathbf{v}\in \mathbb{R}^{q}_{+} \label{eq:dsp_domain}
\end{alignat}\label{eq:BD_original_DSP}
\end{subequations}

\vspace{-15pt}\noindent where $\mathbf{v}$ are the dual variables associated with \mbox{constraints \eqref{eq:sp_constraint}.} Notice that the solution space of the DSP remains the same for different tentative values $\hat{\mathbf{y}}$. 

The MP is given by \eqref{eq:BD_original_RMP}:\vspace{-15pt}

\begin{subequations}
\begin{alignat}{2}
&\!\min_{\mathbf{y,\zeta}}        &\qquad& V_{MP}=\zeta+\mathbf{d}^{T}\mathbf{y}\label{eq:rmp_objfun}\\
&\text{subject to} &               & \mathbf{y}\in \mathbb{Y}^{m} \label{eq:rmp_domain}\\
&                  &               & \zeta \geq \zeta^{low} \label{eq:rmp_zeta_domain}\\
&                  &               & (\mathbf{b}-\mathbf{B}\mathbf{y})^T \mathbf{v}_i \leq \zeta, \ \forall i \in D \label{eq:rmp_optcut}\\
&                  &               & (\mathbf{b}-\mathbf{B}\mathbf{y})^T \mathbf{u}_j \leq 0, \ \forall j \in U \label{eq:rmp_feascut}
\end{alignat}\label{eq:BD_original_RMP}
\end{subequations}

In \eqref{eq:BD_original_RMP} $\zeta$ is a free variable that acts as a surrogate for the value of the DSP. In order to guarantee that the MP is always bounded, an arbitrarily low bound on $\zeta$ is enforced via \eqref{eq:rmp_zeta_domain}. The set of constraints \eqref{eq:rmp_optcut}, where $\mathbf{v}_i$ corresponds to an extreme point of \eqref{eq:BD_original_DSP}, are referred to as optimality Benders cut. If the DSP is unbounded, extreme rays can be extracted instead of extreme points, giving rise to feasibility Benders cuts expressed by \eqref{eq:rmp_feascut}. The sets of extreme points and extreme rays of the DSP are denoted by $D$ and $U$, respectively. 

In each iteration the value of the MP provides a lower bound of \eqref{eq:original_objfun}, i.e., $LB=\hat{\mathbf{\zeta}}+\mathbf{d}^{T}\hat{\mathbf{y}}$. The lower bound is monotonically increasing. A feasible solution of the DSP provides a valid upper bound $UB=(\mathbf{b}-\mathbf{B}\hat{\mathbf{y}})^T \hat{\mathbf{v}}+\mathbf{d}^{T}\hat{\mathbf{y}}$. The algorithm converges if $(UB-LB)<\epsilon$.

Although BD is one of the most widely applied decomposition techniques, the straightforward implementation that was presented in this section is often inefficient from a computational point of view. Several reasons, including poor feasibility and optimality cuts, initial iterations that slowly improve the bounds and slow convergence towards the end of the algorithm, have been recognized. Details about these problems, as well as a systematic review of the research that has been devoted to accelerating BD, can be found in \cite{Rahmaniani_2017}.


\section{Methodology} \label{lab:methodology}
\subsection{HQC Multi-cut Benders Decomposition}

\begin{figure}
    \centering
    \includegraphics[scale=1]{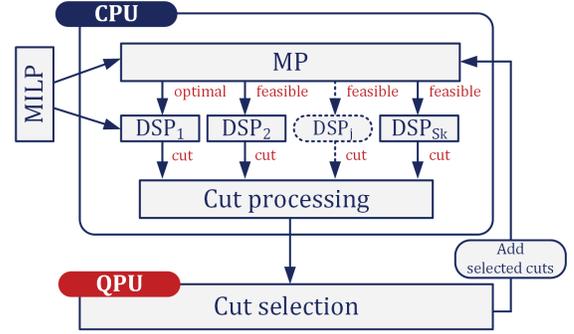}
    \caption{Schematic representation of HQC-MCMS.}
    \label{fig:schematic}
\end{figure}

In this paper, the BD acceleration strategy that was proposed in \cite{Beheshti_2019} and is based on generating multiple cuts via multiple solutions (MCMS) of the MP is explored further. In particular, multiple feasibility and optimality cuts can be generated by exploiting multiple feasible (but not necessarily optimal) solutions of the MP that are available in each iteration. Although all the generated cuts may be appended to the MP in order to reduce the number of iterations, adding a large number of cuts rapidly increases the size of the MP problem and therefore, the total execution time of the algorithm. To achieve a trade-off between the increase in the size of the MP and the reduction in the number of major iterations of the decomposition algorithm, it is possible to add a subset of the cuts that are generated in each iteration. A cut selection subroutine based on pure binary problems that may be solved using QC is developed, hence the modified algorithm is an HQC optimization algorithm (HQC-MCMS). A schematic illustration of HQC-MCMS is shown in Fig. \ref{fig:schematic}. \looseness=-1 

The HQC-MCMS procedure is described in Algorithm \ref{algorithm:MCMS_hybrid}. Similar to the original BD algorithm, the procedure begins by setting the upper and lower bounds of the original objective function to $+\infty$ and $-\infty$, respectively (lines 1 and 2). In each iteration $k$, and while the difference between the upper and lower bounds remains higher than a pre-specified tolerance $\epsilon$, a number of steps are executed. First, the MP is solved and $S_k$ feasible solutions are obtained. Modern solvers permit the extraction of high quality feasible solutions that are found while attempting to solve a MILP problem to optimality. Note that it may be the case that $S_k<S$, where $S$ is the number of requested solutions. If the MP is proven to be infeasible, the optimization problem is also infeasible and the algorithm terminates. Otherwise, the lower bound is updated with the value $V_{MP}^1$ of the optimal solution (lines 5-9).

For each of the $S_k$ available solutions the corresponding DSP is solved. Note that the DSP instances are independent and may be solved in parallel. If a DSP instance is infeasible, the algorithm terminates. If the DSP instance is unbounded, then a feasibility cut is generated and appended to the set of feasibility cuts $G^F_k$, whereas, if the DSP instance is optimal, then an optimality cut is generated and appended to the set of optimality cuts $G^O_k$ (lines 10-19).

The crucial step of the HQC-MCMS algorithm is the cut selection procedure which is described by \mbox{Algorithm 2} (Section \ref{cut_selection_procedure_Algo}) and comprises two elements, namely the cut selection criterion (Section \ref{section:cut_selection_criteria}) and the cut selection strategy (Section \ref{section:cut_selection_strategies}). This is the step where QC can be exploited. After the execution of the cut selection subroutine, the sets of the selected feasibility and optimality cuts, $G^{'F}_k$ and $G^{'O}_k$ respectively, are added to the MP to be solved in the next iteration (lines 20-27).

Finally, the upper bound is updated using the value of the DSP corresponding to the optimal solution of the MP, $V_{DSP}^1$, and the iteration counter is increased. The convergence of the HQC-MCMS algorithm is discussed in \mbox{Section \ref{Convergence}}.

\begin{algorithm}\label{algorithm:MCMS_hybrid}
\caption{HQC-MCMS algorithm}
\KwData{$\epsilon$ (tolerance), $S$ (maximum number of MP solutions to be extracted)}
$UB \gets +\infty$ \;
$LB \gets -\infty$ \;
$k \gets 1$ (iteration counter) \;
\While{$(UB-LB) > \epsilon$}{
    \textbf{Solve} MP$_{k}$ \eqref{eq:BD_original_RMP} and obtain $S_k \leq S$ solutions\; 
    \eIf{infeasible}{
        Stop; Declare infeasibility\;
    }{
        $LB \gets V_{MP}^{1}$ \;
        \For(\Comment*[f]{In parallel}){$j\gets 1$ \KwTo $S_k$ }{ 
            Solve DSP$_{j, k+1}$ \eqref{eq:BD_original_DSP} for $S_j$ \;
            \uIf{infeasible}{
                Stop; Declare infeasibility\;
            }
            \uElseIf{unbounded}{
                Add cut to $G_k^{F}$ \;
            }
            \Else{
                Add cut to $G_k^{O}$ \;
            }
    }
    \If{$G_k^{F} \neq \emptyset$}{
    \textbf{Execute} Algorithm 2 on a \textbf{QPU}\;
    Add selected feasibility cuts $G_k^{'F}$ to MP$_{k+1}$\; 
    }
    \If{$G_k^{O} \neq \emptyset$}{
    \textbf{Execute} Algorithm 2 on a \textbf{QPU}\; 
    Add selected optimality cuts $G_k^{'O}$ to MP$_{k+1}$\; 
    }
    $UB \gets \min\{UB, V_{DSP}^1\}$\;
    }
    $k \gets k+1$\;
}
\end{algorithm}

\subsection{Cut Selection Criteria}\label{section:cut_selection_criteria}

\subsubsection{Criterion I -- cut selection based on the exclusion of infeasible solutions}

In \cite{Beheshti_2019} cut selection was based on the observation that a subset of the generated feasibility cuts may exclude all the infeasible solutions of the MP that are identified in a given iteration. The feasibility cut that corresponds to the infeasible solution $\hat{\mathbf{y}}^i$ of the MP also excludes the infeasible solution $\hat{\mathbf{y}}^j$ if $(\mathbf{b}-\mathbf{B}\hat{\mathbf{y}}^j)^T \hat{\mathbf{u}}_i > 0$. 

Let $|G_{k}^F|$ denote the cardinality of the set of all feasibility cuts that are generated in iteration $k$. Then, the $|G_{k}^F| \times |G_{k}^F|$ binary indicator matrix $\mathbf{E}$ can be constructed in order to compile information about the infeasible MP solutions that are excluded by each cut in the current iteration. Specifically, $\mathbf{E}_{ij}=1$ if the feasibility cut $i$ excludes the infeasible solution associated with the $j$-th solution of the MP, otherwise $\mathbf{E}_{ij}=0$. Note that the diagonal elements of this matrix are always equal to one because, by definition, a feasibility cut excludes the infeasible MP solution based on which it was generated.

Cut selection based on this criterion presents two potential drawbacks. First, this criterion applies only to feasibility cuts. Second, it is possible that for a given problem instance one or more cuts can exclude all the infeasible MP solutions, rendering the application of this criterion trivial.

\subsubsection{Criterion II -- cut selection based on MP variable coverage}

It is often the case that the generated optimality and feasibility cuts are low-density. This means that the coefficient that corresponds to a decision variable of the MP in a given cut is either zero or near-zero relative to other coefficients and therefore, the contribution of a low-density cut to strengthening the MP tends to be limited \cite{Saharidis_2010}. For this reason, several BD acceleration strategies are based on the idea of either generating high-density Pareto optimal cuts \cite{Tang_2013} or bundles of low-density cuts \cite{Saharidis_2010} such that more MP decision variables are covered. In this paper, instead of focusing on the generation of cuts such that the decision variables of the MP are covered, the cuts that are generated based on multiple solutions of the MP are inspected with the purpose of identifying a subset of feasibility (and/or optimality cuts) such that all or most of the MP decision variables are collectively covered. Note that in this paper a rather strict definition of MP variable coverage is adopted. A decision variable $y_i$ of the MP is said to be covered in a given feasibility cut of the form $\sum_{i} y_i (\mathbf{B}^T \mathbf{u})_{i} \geq \mathbf{b}^T \mathbf{u}$, if for the $i$-th row of the matrix $\mathbf{B}^T \mathbf{u}$ it holds $\vert(\mathbf{B}^T \mathbf{u})_{i} \vert>0$. For optimality cuts a similar definition applies if $\mathbf{u}$ is replaced by $\mathbf{v}$. 


A $|G_{k}^F| \times m$ binary indicator matrix $\mathbf{D}^F$ can be constructed after having identified which MP decision variables are covered by a given feasibility cut in the current iteration. Specifically, $\mathbf{D}_{ij}^{F}=1$ if the $j$-th variable of the MP is covered in cut $i$, otherwise $\mathbf{D}_{ij}^{F}=0$. A similar matrix $\mathbf{D}^{O}$ ($|G_{k}^O| \times m$) may be constructed for optimality cuts, if cut selection is to be applied also to the set of optimality cuts. Note that some columns of $\mathbf{D}^F$ and $\mathbf{D}^{O}$ may be zero, that is, some decision variables may not be covered in any of the generated feasibility or optimality cuts.

Compared to Criterion I, cut selection based on the coverage of MP variables applies both to feasibility and optimality cuts and, therefore, may result in more effective MP size management. Moreover, from an implementation perspective, the construction of matrices $\mathbf{D}^F$ and $\mathbf{D}^O$ requires the evaluation of shorter expressions in comparison with $\mathbf{E}$.

\subsection{Cut Selection Strategies}\label{section:cut_selection_strategies}

After having processed the available cuts according to one of the criteria that were presented in \mbox{Section \ref{section:cut_selection_criteria},} a cut selection strategy must be applied in order to identify $G_k^{'F}$ and/or $G_k^{'O}$. Two such strategies based on the minimum set cover problem and the maximum coverage problem, as well as their solution using a QPU are discussed next.

\subsubsection{Strategy I -- minimum set cover for cut selection}\label{section:strategyI}

The minimum set cover problem can be solved to identify a set of cuts with minimum cardinality. If cut selection is based on Criterion I, then the minimum number of feasibility cuts that exclude all the infeasible solutions in the current iteration will be identified. Similarly, if cut selection is based on \mbox{Criterion II}, then the minimum number of feasibility (optimality) cuts that cover all the MP decision variables that can be covered in the current iteration will be identified.

Given a set of rows matrix $\mathbf{M} \in \mathbb{R}^{\vert I \vert \times \vert J \vert}$, where $I$ is the set of rows and $J$ is the set of columns, the minimum set cover problem is a pure integer problem expressed by \eqref{eq:set_cover}: 

\begin{subequations}
\begin{alignat}{2}
&\!\min_{\boldsymbol{\chi}}        &\qquad&  \sum_{i \in I} \chi_i \label{eq:setcover_objfun}\\
&\text{subject to} &               & \sum_{i \in I} M_{ij} \chi_i \geq 1, \ \forall j \in J \label{eq:setcover_const}\\
& &               & \boldsymbol{\chi} \in \{0,1\}^{I}\label{eq:setcover_bin}
\end{alignat}\label{eq:set_cover}
\end{subequations}

\vspace{-10pt}\noindent where $\chi_i$ is a binary variable that is equal to 1 if the cut $i$ is selected and 0 otherwise. It is reiterated that the cut selection problem has to be solved in each iteration $k$. However, for notational simplicity, the iteration index $k$ is dropped. Depending on the cut selection criterion, the columns of $\mathbf{M}$ represent either infeasible solutions or decision variables of the MP (matrix $\mathbf{M}$ is accordingly replaced by matrices $\mathbf{E}$, $\mathbf{D}^{F}$, $\mathbf{D}^{O}$). Note that when Criterion II is used, there is the possibility that a number of MP decision variables cannot be covered and, therefore, \eqref{eq:setcover_const} should be amended by subtracting a binary slack variable from the right-hand side in order to indicate that the constraint cannot be satisfied for a particular column $j$. The sum of the slack variables should also be penalized in \eqref{eq:setcover_objfun}. However, this can be avoided by inspecting $\mathbf{M}$ and dropping all the columns of zeros. 

The minimum set cover problem is known to be \mbox{NP-hard \cite{GROSSMAN_1997}}, i.e., it is intractable for large $I$ and $J$. For this reason, various heuristics have been proposed in order to obtain approximate solutions. In this paper, the set cover problem applied to cut selection is solved using a QPU. For this reason, \eqref{eq:set_cover} must be recast as a QUBO problem. First, \eqref{eq:setcover_const} is converted to an equality constraint by adding an integer slack variable on the right hand side of the constraint and using binary expansion \cite{Tamura_2021} with $\gamma_j = \ceil{\log_{2} \left( \sum_{i \in I} M_{ij} \right)}, \ \forall j \in J$ ancillary qubits as in \eqref{eq:set_cover_slacks}: \looseness=-1 

\begin{equation}
    \sum_{i \in I} M_{ij} \chi_i = 1 + \sum_{\alpha=0}^{\alpha=\gamma_j-1} 2^{\alpha} s_{\alpha j} , \ \forall j \in J \label{eq:set_cover_slacks}
\end{equation}

The QUBO problem formulation is given by \eqref{eq:set_cover_qubo} where $H$ is the Hamiltonian of the problem, and $\mathcal{P_A}$ and $\mathcal{P_B}_j, \forall j \in J$ are positive penalties that are heuristically determined: 

\begin{subequations}
\begin{align}
    & H = H_A +  H_B \\
    & \text{where} \nonumber \\
    & H_A =  \mathcal{P_A} \sum_{i \in I} \chi_i \\
    & H_B = \sum_{j \in J}\mathcal{P_{B}}_j\left[\sum_{i \in I} \left(M_{ij} \chi_i \right) - 1 - \sum_{\alpha=0}^{\alpha=\gamma_j-1}2^{\alpha}s_{\alpha j} \right]^{2}
\end{align}\label{eq:set_cover_qubo}
\end{subequations}

The maximum number of qubits that are required in order to represent the problem is $\vert I \vert + \vert J \vert \cdot \ceil{\log_{2}\vert I \vert}$. If Criterion I is used then this number translates to $\vert G^F_k \vert (1 + \ceil{\log_2 \vert G^F_k \vert})$. This is the case if each infeasible MP solution is excluded by every feasibility cut. If Criterion II is used, then the maximum number of qubits that are needed is $\vert G_{k}^{F} \vert + m \cdot \ceil{\log_2 \vert G_{k}^{F} \vert}$ if all the variables of the MP are covered by every single cut. A similar expression can be written for Criterion II applied to the set of optimality cuts.

It may be observed that for Criterion I the number of qubits that are necessary in order to represent problem \eqref{eq:set_cover_qubo} is independent of the size of the optimization problem \eqref{eq:MILP_problem}. Instead it depends on the user-provided parameter $S$ and is expected to be larger during the first iterations of the algorithm where mostly feasibility cuts are generated. In addition to that, inspection of the columns of matrix $\mathbf{M}$ can reveal rows (feasibility cuts) that exclude all the infeasible solutions in the current iteration and avoid triggering cut selection. The number of qubits in case Criterion II is applied depends on the number of MP variables and may be particularly large. However, for many practical applications the generated cuts are typically expected to be low-density, which implies that a relatively large number of columns of matrix $\mathbf{M}$ are expected to be dropped because all their entries are zero. In both cases, multiple cuts may exclude the same infeasible solutions or cover the same variables. Therefore, the rows of matrix $\mathbf{M}$ can also be inspected in order to remove duplicates, keeping the row that corresponds to a higher quality MP solution.

\subsubsection{Strategy II -- maximum coverage for cut selection}

The conservativeness of Strategy I may lead to a large number of cuts being added to the MP. To address this problem, the maximum coverage problem can be solved as an alterantive cut selection strategy in order to select at most a predefined number of cuts such that, depending on the cut selection criterion that is applied, either the maximum number of infeasible solutions are excluded or the maximum number of MP decision variables are covered. To the best of the Author's knowledge, maximum coverage has not been used as a cut selection strategy in the context of BD before.

Given a matrix $\mathbf{M} \in \mathbb{R}^{\vert I \vert \times \vert J \vert}$ and a maximum number of cuts to be selected $\mathcal{M}$, the maximum coverage problem \cite{Takabe_2018} is a pure integer problem that is formulated in \eqref{eq:max_cover}:

\begin{subequations}
\begin{alignat}{2}
&\!\max_{\boldsymbol{\chi}, \boldsymbol{\phi}}        &\qquad&  \sum_{j \in J} \phi_j \label{eq:maxcover_objfun}\\
&\text{subject to}   &                                & \sum_{i \in I} \chi_i \leq \mathcal{M} \label{eq:maxcover_const2}\\
& &               & \sum_{i \in I} M_{ij} \chi_i \geq \phi_j, \ \forall j \in J \label{eq:maxcover_const1}\\
& &               & \boldsymbol{\chi} \in \{0,1\}^{I}, \boldsymbol{\phi} \in \{0,1\}^{J} \label{eq:maxcover_bin}
\end{alignat}\label{eq:max_cover}
\end{subequations}

The QUBO problem formulation is given by \eqref{eq:max_cover_qubo} where $H$ is the Hamiltonian of the problem and $\mathcal{P_A}$, $\mathcal{P_B}_j, \forall j \in J$ and $\mathcal{P_C}$ are positive penalties that are heuristically determined. First, \eqref{eq:maxcover_const2} is converted into an equality constraint using integer slack variables on the left-hand side of the constraint and binary expansion introducing $\gamma = \ceil{\log_{2} \left( \mathcal{M} +1\right)}$ ancillary qubits. Note that \eqref{eq:maxcover_const2} can be replaced by an equality constraint if $\mathcal{M} \ll \vert I \vert$, thus avoiding the use of slack variables. Similarly, \eqref{eq:maxcover_const1} can be converted into an equality constraint using 
$\gamma_j = \ceil{\log_{2}( \min(\mathcal{M}, \left( \sum_{i \in I} M_{ij}\right))+1)}, \ \forall j \in J$ 
ancillary qubits. \looseness=-1  \vspace{-10pt}


\begin{subequations}
\begin{align}
    & H = H_A +  H_B + H_C \\
    & \text{where} \nonumber \\
    & H_A =  - \mathcal{P_A}\sum_{j \in J} \phi_j \\
    & H_B = \sum_{j \in J}\mathcal{P_B}_{j}\left[\sum_{i \in I} \left(M_{ij} \chi_i \right) - \phi_{j} - \sum_{\alpha=0}^{\alpha=\gamma_j-1}2^{\alpha}s_{\alpha j} \right]^{2}\\
    & H_C = \mathcal{P_C} \left( \sum_{i} \chi_{i}  - \mathcal{M}  + \sum_{\alpha=0}^{\alpha=\gamma-1}2^{\alpha}s_{\alpha}      \right)^2 \label{qubo_max_cover_eq}
\end{align}\label{eq:max_cover_qubo}
\end{subequations}

The maximum number of qubits that are required in order to represent the problem, assumming that \eqref{qubo_max_cover_eq} is replaced by an equality constraint, is 
$\vert I \vert +\vert J \vert \cdot (1+ \ceil{\log_2(\mathcal{M}+1)})$. If Criterion I is used, this number translates to
$\vert G_{k}^{F}  \vert +\vert G_{k}^{F}  \vert \cdot (1+ \ceil{\log_2(\mathcal{M}+1)})$. This is the case if each infeasible MP solution can be excluded by at least $\mathcal{M}$ cuts. If Criterion II is used, then the maximum number of qubits that are required is $\vert G_{k}^{F}  \vert + m \cdot (1+ \ceil{\log_2(\mathcal{M}+1)})$ in case all the MP variables can be covered by at least $\mathcal{M}$ cuts. A similar expression can be written for Criterion II applied to the set of optimality cuts. Although these numbers may appear to be prohibitively large, in practice, the observations of Section \ref{section:strategyI} hold.

\begin{algorithm}\label{algorithm:MCMS_cut_selection}
\caption{Cut selection procedure}
\KwData{$G_k^{F}$, $G_k^{O}$, \textit{cutSelectionCriterion}, \textit{cutSelectionStrategy}, \textit{optSelect} (boolean; whether to apply cut selection on optimality cuts), hyperparameters}
\If{cutSelectionCriterion = Criterion I}{
    Construct $\mathbf{E}$\;
    Inspect $\mathbf{E}$\;

    \If{cutSelectionStrategy = Strategy I}{
        Solve \eqref{eq:set_cover_qubo} to select feasibility cuts\; 
    }
    \If{cutSelectionStrategy = Strategy II}{
        Solve \eqref{eq:max_cover_qubo} to select feasibility cuts\;
    }

    Return $G_k^{'F}$\;
}

\If{cutSelectionCriterion = Criterion II}{
    Construct $\mathbf{D}^{F}$\;
    Inspect $\mathbf{D}^{F}$\;

    \If{cutSelectionStrategy = Strategy I}{
        Solve \eqref{eq:set_cover_qubo} to select feasibility cuts\; 
    }
    \If{cutSelectionStrategy = Strategy II}{
        Solve \eqref{eq:max_cover_qubo} to select feasibility cuts\;
    }
    \If{optSelect}{
        Construct $\mathbf{D}^{O}$\;
        Inspect $\mathbf{D}^{O}$\;
        \If{cutSelectionStrategy = Strategy I}{
            Solve \eqref{eq:set_cover_qubo} to select optimality cuts\; 
        }
        \If{cutSelectionStrategy = Strategy II}{
            Solve \eqref{eq:max_cover_qubo} to select optimality cuts\;
        }
    }
    Return $G_k^{'F}$ and $G_k^{'O}$\;
    
}
\end{algorithm}

\subsection{Cut Selection Procedure}\label{cut_selection_procedure_Algo}

The cut selection problem that is invoked in lines 21 and 25 of Algorithm \ref{algorithm:MCMS_hybrid} is described by Algorithm \ref{algorithm:MCMS_cut_selection}. If cut selection is based on Criterion I, the rows of matrix $\mathbf{E}$ are inspected in order to identify cuts that exclude the same infeasible solutions (line 3). If such duplicate rows are found, the row corresponding to the cut associated with an MP solution with smaller objective function value is kept, while the rest of the rows are dropped. Then, depending on the cut selection strategy, the corresponding optimization problem is solved (lines 5 and 8) and $G_{k}^{'F}$ is returned. If Criterion II is used, matrices $\mathbf{D}^{F}$ and $\mathbf{D}^{O}$ are inspected depending on whether cut selection is applied both to feasibility and optimality cuts (lines 14 and 23). First, their rows are inspected similarly to the rows of $\mathbf{E}$. Then, since it may not be possible to cover a number of MP variables by any of the cuts (all column entries are zero), the respective columns are dropped. Depending on the cut selection strategy, the corresponding optimization problem is solved (lines 25 and 28) using the modified matrix.

The inspection step may significantly reduce the size of the matrices before the cut selection optimization problems are solved. Additionally, all three matrices are inspected in order to identify whether a single cut can either exclude all the infeasible solutions (Criterion I) or covers all the MP variables that can be covered (Criterion II). If such a cut is found, the cut selection procedure terminates without triggering the solution of an optimization problem and the set of selected feasibility and/or optimality cuts that is returned contains only a single cut. \looseness=-1

\subsection{Convergence of HQC-MCMS}\label{Convergence}
Despite the fact that the quantum step does not necessarily provide an optimal (or even feasible) solution of the cut selection problem, the HQC-MCMS algorithm is guaranteed to always converge to the global optimum of Problem \eqref{eq:MILP_problem}. This is due to the fact that sets $G_{k}^{F}$ and $G_{k}^{O}$ and, therefore, any non-empty subsets $G_{k}^{'F}$ and $G_{k}^{'O}$ contain valid cuts \cite{Beheshti_2019}. Moreover, the MP is solved to optimality at each iteration. Then, the feasibility or optimality cut that is generated at the current iteration using the optimal value of the MP in the DSP can always be added in the next iteration to prevent $G_{k}^{'F}$ and $G_{k}^{'O}$ from being empty. Note that according to \cite{Geoffrion_1974}, it is also possible to avoid solving the MP to optimality in each iteration while guaranteeing convergence; however, the practical implications are out of the scope of this study.

\begin{table}[t]
\centering
\caption{Notation Used in the UC Formulation\vspace{-5pt}}\label{UC:Nomenclature}
\begin{tabular}{p{0.07\linewidth}p{0.85\linewidth}}
\toprule
\multicolumn{2}{l}{\textbf{Sets and indices}} \\
$g \ (G)$             & Index (set) of generators \\
$t \ (T)$             & Index (set) of time \\
$i,j \ (I)$           & Indices (set) of buses \\
$I^{0}$             & Set of reference bus \\
$l (L)$             & Index (set) of loads \\
&  \\
\multicolumn{2}{l}{\textbf{Parameters}} \\
$A_{ig}^{G}$    & Generator incidence matrix; 1 if generator $g$ is connected to \mbox{bus $i$} \\
$A_{il}^{L}$    & Load incidence matrix; 1 if load $l$ is connected to bus $i$ \\
$C_{g}$         & Energy cost of generator $g$ (\$/MWh) \\
$D_{lt}$         & Demand of load $l$ in time $t$ (MW) \\
$NLC_{g}$       & No-load cost of generator $g$ (\$/h) \\
$SUC_{g}$       & Start-up cost of generator $g$ (\$)\\
$SDC_{g}$       & Shut-down cost of generator $g$ (\$)\\
$P_{g}^{max}$   & Maximum power output of generator $g$ (MW)\\
$P_{g}^{min}$   & Minimum power output of generator $g$ (MW)\\
$P_{g}^{ini}$   & Initial power output of generator $g$ (MW)\\
$RU_{g}$        & Ramp-up rate of generator $g$ (MW/h)\\
$RD_{g}$        & Ramp-down rate of generator $g$ (MW/h)\\
$X_{ij}$        & Reactance of line $(i,j)$ (pu) \\
$F_{ij}^{max}$  & Maximum power flow through line $(i,j)$ (MW) \\
$B_{ij}$        & $(i,j)$ admittance matrix element (pu) \\
$u_{g}^{ini}$   & Initial state of generator $g$; 1 if generator $g$ was online before the beginning of the scheduling horizon \\
$\zeta^{low}$   & Lower bound of the proxy variable $\zeta$ (\$) \\
                &  \\
\multicolumn{2}{l}{\textbf{Decision variables}} \\
$P_{gt}$        & Power output of generator $g$ in time $t$ (MW) \\
$\theta_{it}$   & Voltage angle of bus $i$ in time $t$ (rad) \\ 
$\zeta$         & Proxy for the DSP value (\$) \\
$u_{gt}$        & Binary variable; 1 if generator $g$ is online in time $t$\\
$y_{gt}$        & Binary variable; 1 if generator $g$ is starting-up in time $t$ \\
$z_{gt}$        & Binary variable; 1 if generator $g$ is shutting-down in time $t$ \\
\bottomrule
\end{tabular}
\end{table}

\section{Use Case: The Unit Commitment Problem}\label{lab:use_cases}

The UC problem is a fundamental power system optimization problem that aims to optimaly schedule and dispatch the available generation and demand side resources \cite{Zheng_2015}. Various BD schemes have been applied in order to solve different versions of this problem \cite{Wen_2016, Nasri_2016, Alemany_2013, Yong_2013, Fu_2005}. To demonstrate the applicability of the proposed HQC algorithm a simple version of the UC problem with typical operational and network constraints is adopted. First, in Section \ref{lab:use_cases_formulation} the complete MILP problem formulation is presented. Then, the DSP and MP formulations are derived in Sections \ref{lab:use_cases_dsp} and \ref{lab:use_cases_rmp}.

\subsection{MILP Formulation}\label{lab:use_cases_formulation}
The UC problem formulation is described by \eqref{eq:UCor_obj}-\eqref{eq:UCor_dom3}. The notation that is used is presented in Table \ref{UC:Nomenclature}. The dual variables associated with each constraint set are denoted by greek letters that are displayed in parentheses next to the corresponding equation. Note that all the dual variables associated with the inequality constraints are non-negative, whereas the dual variables associated with equality constraints are unrestricted.  \looseness=-1

\begin{subequations}
\begin{align}
    &\min_{\boldsymbol{P},\boldsymbol{\theta}, \boldsymbol{u},\boldsymbol{y},\boldsymbol{z}} \quad TC \nonumber\\
    &\text{where} \ TC = \sum_{t}\sum_{g} \left( C_{g} P_{gt}  \right. \nonumber \\
    & \left. \hspace{55pt}+ NLC_{g} u_{gt} +SUC_{g}y_{gt}+SDC_{g}z_{gt} \right)  \label{eq:UCor_obj}\\
    &\text{subject to:} \nonumber \\ 
    & y_{gt}-z_{gt} = u_{gt}-u_{g(t-1)}                 \ \ \ \forall g, t>1 \label{eq:UCor_com1}\\
    & y_{gt}-z_{gt} = u_{gt}-u_{g}^{ini}                \ \ \ \forall g, t=1 \label{eq:UCor_com2}\\
    & -P_{gt} \geq -P_{g}^{max} u_{gt}     \ \ \ \forall g,t \ \ (\mu_{gt}^{+})     \label{eq:Pmax}\\
    & P_{gt}  \geq P_{g}^{min} u_{gt}       \ \ \ \forall g,t \ \ (\mu_{gt}^{-})    \label{eq:Pmin}\\
    & -P_{gt} + P_{g(t-1)} \geq -RU_{g}                 \ \ \ \forall g,t>1 \ \ (\nu^{+}_{gt})      \label{eq:ramp1}\\
    & -P_{gt}  \geq -RU_{g}-P_{g}^{ini}                 \ \ \ \forall g,t=1 \ \ (\nu^{0+}_{g})      \label{eq:ramp2}\\
    & P_{gt} - P_{g(t-1)} \geq - RD_{g}                 \ \ \ \forall g,t>1 \ \ (\nu^{-}_{gt})      \label{eq:ramp3}\\
    & P_{gt} \geq - RD_{g} + P_{g}^{ini}                \ \ \ \forall g,t=1 \ \ (\nu^{0-}_{g})      \label{eq:ramp4}\\
    & \frac{1}{X_{ij}} (\theta_{jt}-\theta_{it}) \geq - F_{ij}^{max}     \ \ \ \forall (i,j)|X_{ij}\neq0,t \ \ (\psi_{ijt})     \label{eq:power_flow}\\
    & \theta_{i,t} = 0 \ \ \ \forall t, i\in I^0 \ \ (\lambda^{0}_{t})  \label{eq:reference_bus}\\
    & \sum_{j \in I}B_{ij}\theta_{jt} - \sum_{g\in G} P_{gt}A_{ig}^{G} = - \sum_{l \in L}D_{lt}A_{il}^{L} \ \  \forall i,t \ \ (\lambda_{it}) \label{eq:bus_balance}\\
    & P_{gt} \geq 0 \ \ \ \forall g,t                           \label{eq:UCor_dom1}\\ 
    & \theta_{it} \in \mathbb{R} \ \ \ \forall i,t              \label{eq:UCor_dom2}\\
    & u_{gt}, y_{gt}, z_{gt} \in \{0,1\} \ \ \ \forall g,t      \label{eq:UCor_dom3}
\end{align}
\end{subequations}

The objective function is expressed by \eqref{eq:UCor_obj} and stands for the minimization of the total energy and commitment cost across the time horizon. Constraints \eqref{eq:UCor_com1} and \eqref{eq:UCor_com2} determine the generator commitment logic. 
The power output of generators is limited by \eqref{eq:Pmax} and \eqref{eq:Pmin}. The intertermporal constraints \eqref{eq:ramp1}-\eqref{eq:ramp4} limit the change in the power output of generators in consecutive time periods according to their up and down ramp rates. Network constraints are modeled in terms of a DC power flow approximation. The power that flows through transmission lines is constrained by \eqref{eq:power_flow}, while \eqref{eq:reference_bus} fixes the voltage angle at the reference bus. The power balance at each bus is determined by \eqref{eq:bus_balance}. Finally, \eqref{eq:UCor_dom1}-\eqref{eq:UCor_dom3} determine the domain of the decision variables.

\subsection{Dual Subproblem}\label{lab:use_cases_dsp}

For a tentative solution $\hat{u}_{gt}$ of the MP, the DSP is an LP problem that is expressed by \eqref{eq:UC_DSP_objective}-\eqref{eq:UC_DSP_last_constraint}.

\begin{subequations}
\begin{align}
    & \max_{\boldsymbol{\mu}^{+}, \boldsymbol{\mu}^{-}, \boldsymbol{\nu}^{+}, \boldsymbol{\nu}^{-}, \boldsymbol{\nu}^{0-}, \boldsymbol{\nu}^{0+}, \boldsymbol{\psi}, \boldsymbol{\lambda}, \boldsymbol{\lambda^{0}}} \quad UC^{DSP} \nonumber\\
    & \text{where} \ UC^{DSP}= \sum_{t}\sum_{g} \left( \mu_{gt}^{-} P_{g}^{min} -\mu_{gt}^{+}P_{g}^{max} \right) \hat{u}_{gt}  \nonumber      \\
    & -\sum_{t}\sum_{g} \left( \nu_{gt}^{+} RU_{g}   + \nu_{gt}^{-} RD_{g}  \right)                   \nonumber      \\
    & +\sum_{g} \left[ \nu_{g}^{0+}\left(-P_{g}^{ini}-RU_{g} \right)  + \nu_{g}^{0-} \left(P_{g}^{ini}-RD_{g} \right)  \right]              \nonumber \\
    & -\sum_{t}\sum_{i} \left[\lambda_{it} \sum_{l}\left(D_{lt}A_{il}^{L}  \right) +\sum_{j |X_{ij}\neq 0} \psi_{ijt}F_{ij}^{max}  \right]     \label{eq:UC_DSP_objective}\\
    & \nonumber \\
    & \text{subject to:} \nonumber \\
    & -\mu_{gt}^{+} + \mu_{gt}^{-}-\nu_{g}^{0+}+\nu_{g}^{0-}+\nu_{g(t+1)}^{+}-\nu_{g(t+1)}^{-} \nonumber\\
    & \hspace{65pt} -\sum_{i} A_{ig}^{G}\lambda_{it} \leq C_{g}   \ \ \ \forall g,t=1 \\ 
    & -\mu_{gt}^{+} + \mu_{gt}^{-}-\nu_{gt}^{+}+\nu_{gt}^{-} \nonumber \\
    & \hspace{65pt} -\sum_{i} A_{ig}^{G}\lambda_{it} \leq C_{g} \ \ \ \forall g,t=|T| \\
    & -\mu_{gt}^{+} + \mu_{gt}^{-} - \nu_{gt}^{+} + \nu_{g(t+1)}^{+} + \nu_{gt}^{-} - \nu_{g(t+1)}^{-} \nonumber \\
    & \hspace{65pt} -\sum_{i} A_{ig}^{G}\lambda_{it} \leq C_{g} \ \ \ \forall g,t\in(1,|T|) \\
    & \lambda_{t}^{0} + \sum_{j} \left( \lambda_{jt}B_{ji} \right) \nonumber \\
    & \hspace{45pt} +\sum_{j|X_{ij}\neq0} \left(\frac{\psi_{jit}-\psi_{ijt}}{X_{ij}} \right) = 0 \ \ \ \forall i \in I^{0},t \\ 
    & \sum_{j} \left( \lambda_{jt}B_{ji} \right) \nonumber\\
    & \hspace{30pt} +\sum_{j|X_{ij}\neq0} \left(\frac{\psi_{jit}-\psi_{ijt}}{X_{ij}} \right) = 0 \ \ \ \forall i \in I-I^{0},t \\
    & \mu_{gt}^{+}, \mu_{gt}^{-}, \nu_{gt}^{+}, \nu_{gt}^{-} \geq 0 \ \ \ \forall g,t \\
    & \nu_{g}^{0-}, \nu_{g}^{0+} \geq 0 \ \ \ \forall g \\
    & \psi_{ijt} \geq 0 \ \ \ \forall (i,j)|X_{ij} \neq 0, t \\
    & \lambda_{t}^{0} \in \mathbb{R} \ \ \ \forall t \\
    & \lambda_{it} \in \mathbb{R} \ \ \ \forall i,t \label{eq:UC_DSP_last_constraint}
\end{align}
\end{subequations}

Constraints involving binary variables (e.g., generator minimum up and down time, reserve capacity constraints, etc.) can be directly added to the formulation of Section \ref{lab:use_cases_formulation} since they do not affect the DSP.

\subsection{Master Problem}\label{lab:use_cases_rmp}

Given $G_{k}^{'F}$ and $G_{k}^{'O}$, the MP is a MILP problem that is expressed by \eqref{eq:UCRMP_obj}-\eqref{eq:UCRMP_last_constraint}. 

\begin{subequations}
\begin{align}
    & \min_{\zeta, \boldsymbol{u},\boldsymbol{y},\boldsymbol{z} } \quad \zeta + \sum_{t}\sum_{g} \left(NLC_{g} u_{gt} +SUC_{g}y_{gt}+SDC_{g}z_{gt}\right) \label{eq:UCRMP_obj}\\
    & \text{subject to:} \nonumber \\
    & \eqref{eq:UCor_com1},\eqref{eq:UCor_com2} \nonumber \\
    & \left\{\sum_{t}\sum_{g} \left( \hat{\mu}_{gt}^{-} P_{g}^{min} -\hat{\mu}_{gt}^{+}P_{g}^{max} \right) u_{gt}  \nonumber \right.     \\
    & -\sum_{t}\sum_{g} \left( \hat{\nu}_{gt}^{+} RU_{g}   + \hat{\nu}_{gt}^{-} RD_{g}  \right)                   \nonumber      \\
    & +\sum_{g} \left[ \hat{\nu}_{g}^{0+}\left(-P_{g}^{ini}-RU_{g} \right)  + \hat{\nu}_{g}^{0-} \left(P_{g}^{ini}-RD_{g} \right)  \right]              \nonumber \\
    &  -\sum_{t}\sum_{i} \left[\hat{\lambda}_{it} \sum_{l}\left(D_{lt}A_{il}^{L}  \right) +\sum_{j |X_{ij}\neq 0} \hat{\psi}_{ijt}F_{ij}^{max}  \right]        \nonumber \\
    &\hspace{45pt}\left. \phantom{\sum_{t}P^{test}} \leq \zeta \right\}^{(\kappa)}   \in \bigcup\limits_{\kappa=2}^{k} G_{\kappa}^{'O}            \label{eq:UCRMP_OPT} \\
    & \nonumber \\
    & \left\{ \sum_{t}\sum_{g} \left( \hat{\mu}_{gt}^{-} P_{g}^{min} -\hat{\mu}_{gt}^{+}P_{g}^{max} \right) u_{gt}  \nonumber  \right.    \\
    & -\sum_{t}\sum_{g} \left( \hat{\nu}_{gt}^{+} RU_{g}   + \hat{\nu}_{gt}^{-} RD_{g}  \right)                   \nonumber      \\
    & +\sum_{g} \left[ \hat{\nu}_{g}^{0+}\left(-P_{g}^{ini}-RU_{g} \right)  + \hat{\nu}_{g}^{0-} \left(P_{g}^{ini}-RD_{g} \right)  \right]              \nonumber \\
    & -\sum_{t}\sum_{i} \left[\hat{\lambda}_{it} \sum_{l}\left(D_{lt}A_{il}^{L}  \right) +\sum_{j |X_{ij}\neq 0} \hat{\psi}_{ijt}F_{ij}^{max}  \right]            \nonumber \\
    &\hspace{45pt} \left.\phantom{\sum_{t}P^{test}} \leq 0 \right\}^{(\kappa)} \in \bigcup\limits_{\kappa=2}^{k} G_{\kappa}^{'F}    \label{eq:UCRMP_FEAS}\\
    & \zeta \geq \zeta^{low}  \label{eq:UCRMP_zlow}\\ 
    & u_{gt}, y_{gt}, z_{gt} \in \left\{0,1\right\} \ \ \ \forall g,t \label{eq:UCRMP_last_constraint}
\end{align}
\end{subequations}

The optimality and feasibility cuts that are appended to the MP up to the current iteration $k$ are expressed by \eqref{eq:UCRMP_OPT} and \eqref{eq:UCRMP_FEAS} respectively. Constraint \eqref{eq:UCRMP_zlow} is necessary in order to prevent the problem from being unbounded in the first iteration of the algorithm. 

\section{Numerical Experiments}\label{lab:numerical_experiments}

\subsection{Implementation Details}

The HQC-MCMS algorithm was implemented in \mbox{Python 3.9} using the Pyomo package \cite{Pyomo_2021}. All the classical MILP problems were solved using the Gurobi 9.5.0 \mbox{solver \cite{Gurobi_2021}} with a \textit{MIPGap} of \mbox{0 \%} on a workstation with 2 Intel Xeon processors (24 cores, 3 GHz) and 128 GB of RAM. In order to find multiple solutions of the MP the solution pool functionality of Gurobi was used. The \textit{PoolSearchMode} parameter was set to 1. This means that the solver continues the search for feasible solutions after the optimal solution has been found, however, no guarantees are provided about the quality of the additional feasible solutions. 

The QUBO problem instances were solved using the \mbox{D-Wave} Advantage 4.1 QA that can be accessed via Amazon Braket. The \mbox{D-Wave} Advantage 4.1 QPU relies on the Pegasus topology and features more than 5000 qubits and 35000 couplers \cite{Advantage_manual}. Embedding of the problem on the physical QPU graph was performed using the \textit{minorminer} package using the default settings. For all problems that were submitted to the QPU the annealing time was set to 10 $\mu$s and the anneal-read cycle was repeated 1000 times. It is to be noted that the hyperparameters associated with solving problems on a QPU may impact the solution time and quality. Thus, they should be carefully tuned. However, due to the considerable utilization cost of QPUs and the diversity of problem instances that are solved, only a limited number of tuning attempts were performed in this study. The chain strength value was set to 150\% of the largest interaction coefficient observed in the problem Hamiltonian. According to the recommendation in \cite{Lucas_2014} for general binary integer linear problems for problems of the type \eqref{eq:set_cover_qubo}, $\mathcal{P_A}=1$ and $\mathcal{P_B}_j = \vert I \vert, \ \forall j \in J$ and for problems of the type \eqref{eq:max_cover_qubo}, $\mathcal{P_A}=1$, $\mathcal{P_B}_j = \vert I \vert+ \vert J\vert, \ \forall j \in J$ and $\mathcal{P_C}= \vert I\vert +\vert J \vert $, where $I,J$ are the sets of rows and columns of matrix $\mathbf{M}$ respectively.

\begin{figure}[t] 
\centering
  \subfigure[8-bus power system with 13 transmission lines and 1071 MW generating capacity]{\includegraphics[scale=0.85]{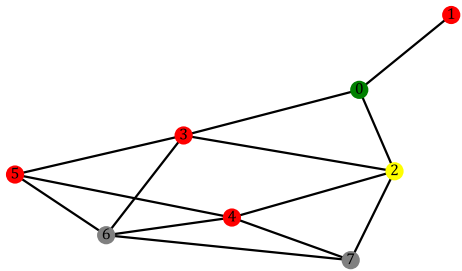}} 
  \subfigure[30-bus power system with 51 transmission lines and 5395 MW generating capacity]{\includegraphics[scale=0.85]{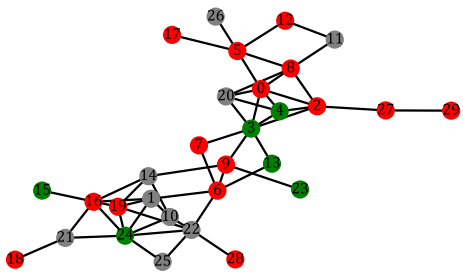}}
  \caption{The test power systems. The nodes are color-coded. Red: load bus, Green: generation bus, Gray: load and generation bus, Yellow: transfer bus. Without loss of generality, Bus 0 is defined as the reference bus.}
  \label{fig:test_systems} 
\end{figure}

 
\begin{figure}
    \centering
    \includegraphics[scale=0.85]{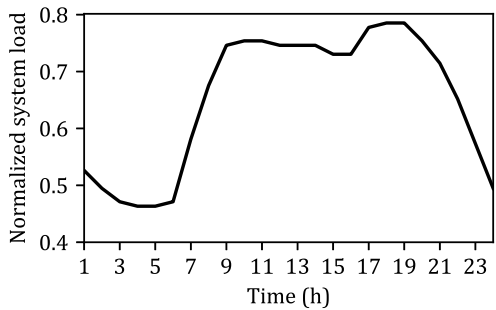}
    \caption{Normalized system load. Extracted from \cite{IEEE_24_DTU}.}
    \label{fig:system_load}
\end{figure}

\subsection{Input Data}

To investigate the applicability of the proposed methodology on the UC problem, the test systems displayed in Fig. \ref{fig:test_systems} were randomly generated. The network topologies were generated based on the methodology that was proposed in \cite{Wang_2008}. First, buses are uniformly placed within a fixed area with width and height of 1. Then, given a distance requirement (in this study $[0,0.4]$) between neighboring buses, the set of transmission lines are selected by sampling a Poisson distribution with its parameter set to $2.67$. The reactance of a transmission line is considered proportional to its length. For simplicity a factor of 1/10 is assumed for all lines. Subsequently, the type of each bus is decided. It is assumed that 50\% of the buses are load buses, 20\% are generator buses and 30\% have both loads and generators connected. In case the number of buses is such that the aforementioned percentages do not result in integers with a sum equal to the number of buses, the remaining buses are considered to be transfer buses (i.e., they do not connect loads or generators). The capacity of each line is sampled from a uniform distribution ranging from 15\% to 35\% of the total generating capacity of the system. The individual loads are assigned a percentage of the hourly normalized system load (with respect to the maximum generating capacity of the system) that is portrayed in \mbox{Fig. \ref{fig:system_load}} for 24 time periods using the Dirichlet distribution. The Dirichlet distribution satisfies the requirements that each bus load fraction is positive and that the sum of the fractions is 1. Finally, generator parameters are constructed using the values presented in Table \ref{UC:generator_parameters}.

\begin{table}[t]
\centering
\caption{Generator Parameters. Ranges Imply Sampling a Value From a Uniform Distribution}\label{UC:generator_parameters}
\begin{tabular}{p{0.1\linewidth}p{0.1\linewidth}p{0.6\linewidth}}
\toprule
\textbf{Parameter}  &  & \textbf{Value} \\
\midrule
$P_{g}^{max}$   & & $[60,600]$\\
$P_{g}^{min}$   & & $[20\% P_{g}^{max},40\% P_{g}^{max}]$\\
$RU_{g}$        & & $\max(P_{g}^{min},[20\%P_{g}^{max},40\%P_{g}^{max}])$\\
$RD_{g}$        & & $RU_{g}$\\
$SUC_{g}$       & & $[5,1600]$\\
$SDC_{g}$       & & $SUC_{g}$\\
$C_{g}$         & & $[5,30]$ \\
$NLC_{g}$       & & $[3 C_{g},6 C_{g}]$ \\
$u_{g}^{ini}$   & & $\left\{0,1\right\}$ \\
$P_{g}^{ini}$   & & $P_{g}^{min} u_{g}^{ini}$ \\
\bottomrule
\end{tabular}
\end{table}

\begin{table*}[t]
  \centering
  \caption{Solution Times for the 8-bus System. All Times Are in (s)}
  \setlength\tabcolsep{3pt}
    \begin{tabular}{p{8.665em}ccc|ccc|ccc|ccc|ccc}
    \toprule
    \multicolumn{1}{r}{} &   &   & \multicolumn{1}{r}{} & \multicolumn{6}{|c|}{\textbf{Classical}} & \multicolumn{6}{c}{\textbf{Quantum}} \\
    \midrule
    \multicolumn{1}{r}{} & \textbf{BD} & \textbf{All cuts} & \textbf{Random} & \textbf{C1} & \textbf{C2} & \textbf{C3} & \textbf{C4} & \textbf{C5} & \textbf{C6} & \textbf{C7} & \textbf{C8} & \textbf{C9} & \textbf{C10} & \textbf{C11} & \textbf{C12} \\
    \midrule
    \textit{cutSelectionStrategy} & - & - & - & \multicolumn{3}{c|}{I} & \multicolumn{3}{c|}{II} & \multicolumn{3}{c|}{I} & \multicolumn{3}{c}{II} \\
    \midrule
    \textit{cutSelectionCriterion} & - & - & - & I & II & II & I & II & II & I & II & II & I & II & II \\
    \midrule
    \textit{optSelect} & - & - & - & - & F & T & - & F & T & - & F & T & - & F & T \\
    \midrule
    \textbf{Iterations} & 23.00 & 7.00 & 17.60 & 13.00 & 7.00 & 7.00 & 13.00 & 9.00 & 9.00 & 13.00 & 7.00 & 7.00 & 13.40 & 10.00 & 9.80 \\
    \textbf{Time} & 28.20 & 8.66 & 19.10 & 13.12 & 9.22 & 9.24 & 12.95 & 10.65 & 10.78 & 13.14 & 9.06 & 10.33 & 14.87 & 17.22 & 64.62 \\
    \textbf{MP solution} & 1.74 & 2.17 & 3.77 & 1.77 & 1.63 & 1.30 & 1.71 & 1.86 & 1.75 & 1.76 & 1.58 & 1.39 & 1.89 & 2.10 & 1.68 \\
    \textbf{DSP solution} & 26.46 & 6.49 & 15.33 & 8.36 & 6.50 & 6.61 & 8.34 & 7.15 & 7.13 & 8.37 & 6.58 & 6.63 & 8.54 & 7.55 & 7.47 \\
    $\mathbf{M}$ \textbf{construction} & - & - & - & 2.37 & 0.32 & 0.42 & 2.29 & 0.49 & 0.54 & 2.27 & 0.33 & 0.40 & 2.30 & 0.52 & 0.58 \\
    \textbf{Cut selection} & - & - & - & 0.62 & 0.78 & 0.92 & 0.62 & 1.16 & 1.35 & 0.33 & 0.39 & 0.50 & 0.43 & 0.82 & 0.94 \\
    \multicolumn{1}{l}{\textbf{Minor-emb. (feas.)}} & - & - & - & - & - & - & - & - & - & 0.41 & 0.17 & 0.18 & 1.71 & 6.23 & 7.62 \\
    \textbf{Minor-emb. (opt.)} & - & - & - & - & - & - & - & - & - & - & - & 1.23 & - & - & 46.34 \\
    \bottomrule
    \multicolumn{16}{r}{T: True, F: False}
    \end{tabular}%
  \label{tab:results_8bus}%
\end{table*}%

\begin{table*}[t]
  \centering
  \caption{Solution Times for the 30-bus System. All Times Are in (s)}
    \setlength\tabcolsep{3pt}
    \begin{tabular}{p{8.665em}ccc|ccc|ccc|ccc|ccc}
    \toprule
    \multicolumn{1}{r}{} &   &   &   & \multicolumn{6}{c|}{\textbf{Classical}} & \multicolumn{6}{c}{\textbf{Quantum}} \\
    \midrule
    \multicolumn{1}{r}{} & \multicolumn{1}{p{2.89em}}{\textbf{BD}} & \textbf{All cuts} & \textbf{Random} & \textbf{C1} & \textbf{C2} & \textbf{C3} & \textbf{C4} & \textbf{C5} & \textbf{C6} & \textbf{C7} & \textbf{C8} & \textbf{C9} & \textbf{C10} & \textbf{C11} & \textbf{C12} \\
    \midrule
    \textit{cutSelectionStrategy} & - & - & - & \multicolumn{3}{c|}{I} & \multicolumn{3}{c|}{II} & \multicolumn{3}{c|}{I} & \multicolumn{3}{c}{II} \\
    \midrule
    \textit{cutSelectionCriterion} & - & - & - & I & II & II & I & II & II & I & II & II & I & II & II \\
    \midrule
    \textit{optSelect} & - & - & - & - & F & T & - & F & T & - & F & T & - & F & T \\
    \midrule
    \textbf{Iterations} & 51.00 & 10.00 & 24.40 & 9.00 & 11.00 & 10.00 & 9.00 & 9.00 & 9.00 & 9.00 & 10.60 & 8.40 & 9.00 & 10.20 & 11.40 \\
    \textbf{Time} & 110.50 & 80.56 & 69.65 & 53.22 & 64.73 & 35.01 & 52.86 & 38.11 & 29.16 & 56.31 & 109.92 & 1348.55 & 73.14 & 223.53 & 1594.90 \\
    \textbf{MP solution} & 29.00 & 62.86 & 37.58 & 27.74 & 40.67 & 12.25 & 26.51 & 19.19 & 8.71 & 28.40 & 45.88 & 18.04 & 23.33 & 30.20 & 12.24 \\
    \textbf{DSP solution} & 81.51 & 17.70 & 32.07 & 16.12 & 20.45 & 17.82 & 16.10 & 16.17 & 15.83 & 15.95 & 18.77 & 19.07 & 16.29 & 17.91 & 23.40 \\
    $\mathbf{M}$ \textbf{construction} & - & - & - & 8.77 & 1.92 & 2.56 & 9.64 & 1.74 & 2.31 & 8.91 & 1.63 & 2.11 & 9.60 & 1.84 & 3.04 \\
    \textbf{Cut selection} & - & - & - & 0.59 & 1.70 & 2.38 & 0.61 & 1.00 & 2.30 & 0.37 & 1.02 & 1.91 & 0.44 & 1.08 & 2.61 \\
    \multicolumn{1}{l}{\textbf{Minor-emb. (feas.)}} & - & - & - & - & - & - & - & - & - & 2.67 & 42.62 & 53.43 & 23.49 & 172.51 & 360.04 \\
    \textbf{Minor-emb. (opt.)} & - & - & - & - & - & - & - & - & - & - & - & 1253.99 & - & - & 1193.57 \\
    \bottomrule
    \multicolumn{16}{r}{T: True, F: False}
    \end{tabular}%
  \label{tab:results_30bus}%
\end{table*}%

\subsection{Results and Discussion}

In this section, the results of the numerical experiments are reported. For all the experiments the algortihm terminates if $\frac{UB-LB}{UB}100\%<0.5\%$. For the 8-bus system, the whole 24-hour load profile is used (72 MP decision variables are involved in the DSP), while 10 solutions of the MP are requested from the solver. For the 30-bus system 30 solutions of the MP are requested, however, considering the full 24 periods the heuristic would not result in finding a feasible minor-embedding within the timeout limit of 1000s for all the cases. For this reason, only the first 8 time periods were considered (120 MP decision variables variables involved in the DSP). For both power systems, Strategy II was applied using $\mathcal{M}=3$ (both for feasibility and optimality cuts). It should be noted that the lowest-energy solution returned by the QPU did not always satisfy the constraints of the cut selection strategy. In these cases, the lowest-energy valid solution was kept. If none of the solutions in the returned sample set were valid, the solution with the lowest number of constraint violations was implemented.

\subsubsection{Performance comparison of cut selection criteria and strategies}

In order to assess the performance of the HQC-MCMS algorithm the cut selection procedure described in Algorithm 2 is executed both on a classical computer and a QA. Detailed computational characteristics for the two test systems are presented in Tables \ref{tab:results_8bus} and \ref{tab:results_30bus}. The values presented in these tables are averaged over 5 executions of the algorithm for each case. The straightforward implementation of BD and the addition of all the available cuts to the MP are used as benchmarks of the performance of the multi-cut strategies. Moreover, a case in which the optimal solution as well as three randomly selected solutions of the MP are used to generate cuts is presented in order to establish the fact that Algorithm 2 performs a non-trivial cut selection. For ease of reference, the different combinations of cut selection strategies and cut selection criteria are denoted by C1-C12. The time for completing a quantum computation is given by $T_{Q} \approx T_{P} + \rho(T_{A}+ T_{R})$, where $T_{Q}$ is the QPU access time, $T_{P}$ is the quantum programming time, $\rho$ is the number of times an anneal-read cycle is repeated, $T_{A}$ is the annealing time and $T_{R}$ is the time required to read a measurement. It is conventional to report $\rho T_{A}$ as an equivalent to the classical CPU time \cite{Jones_2020}, however, $T_Q$ is reported in Tables \ref{tab:results_8bus} and \ref{tab:results_30bus} for the cut selection component for the sake of completeness. 

\begin{figure*}[t] 
\centering
  \subfigure[Strategy I, Criterion I (C7)]{\includegraphics[scale=0.7]{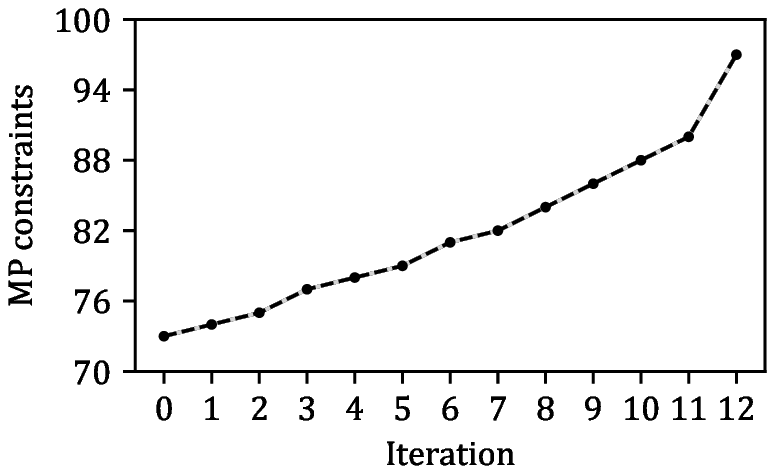}}
  \subfigure[Strategy I, Criterion II, \textit{optSelect}=False (C8)]{\includegraphics[scale=0.7]{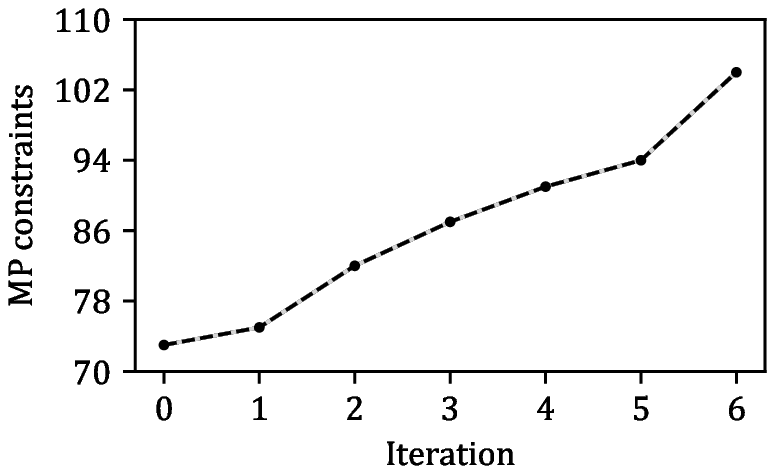}}
  \subfigure[Strategy I, Criterion II, \textit{optSelect}=True (C9)]{\includegraphics[scale=0.7]{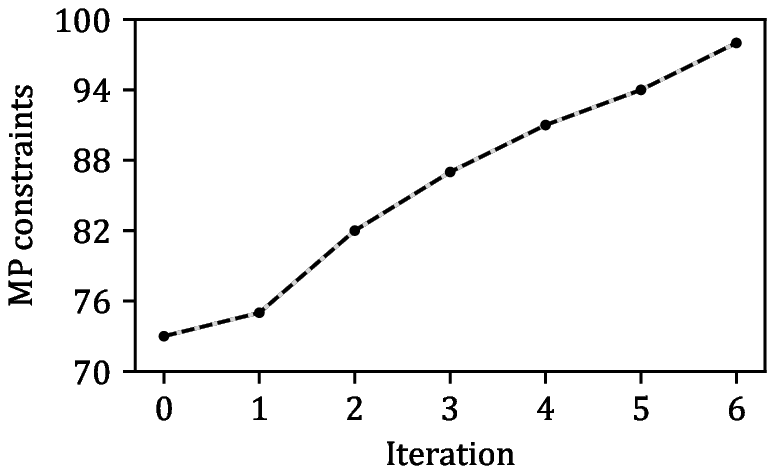}}
  \subfigure[Strategy II,Criterion I (C10)]{\includegraphics[scale=0.7]{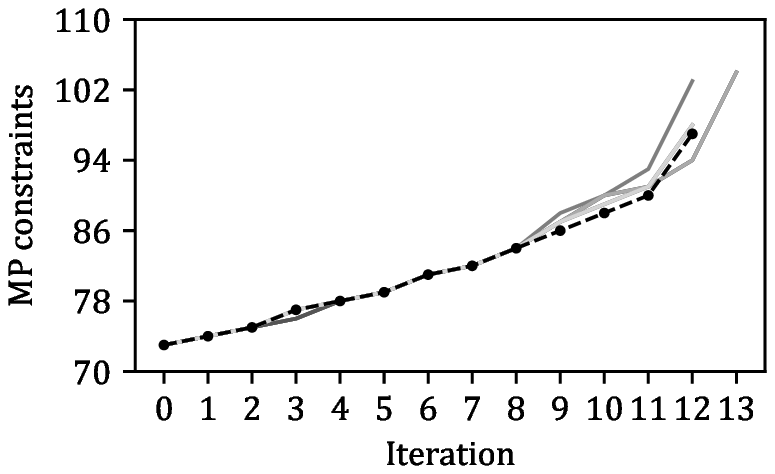}}
  \subfigure[Strategy II, Criterion II, \textit{optSelect}=False (C11)]{\includegraphics[scale=0.7]{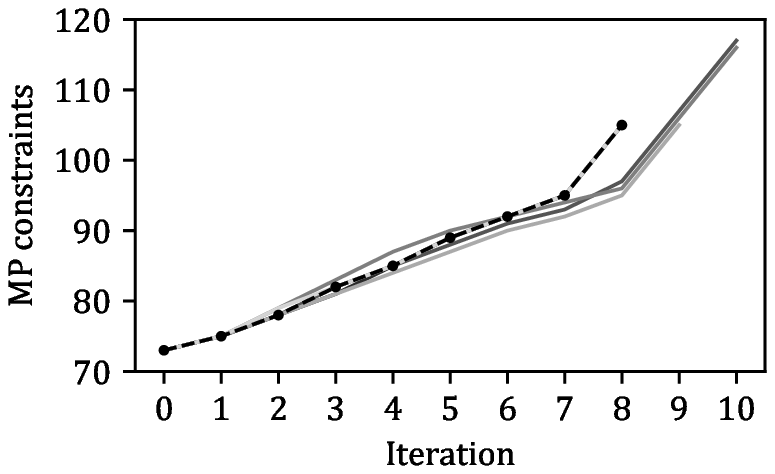}}
  \subfigure[Strategy II, Criterion II, \textit{optSelect}=True (C12)]{\includegraphics[scale=0.7]{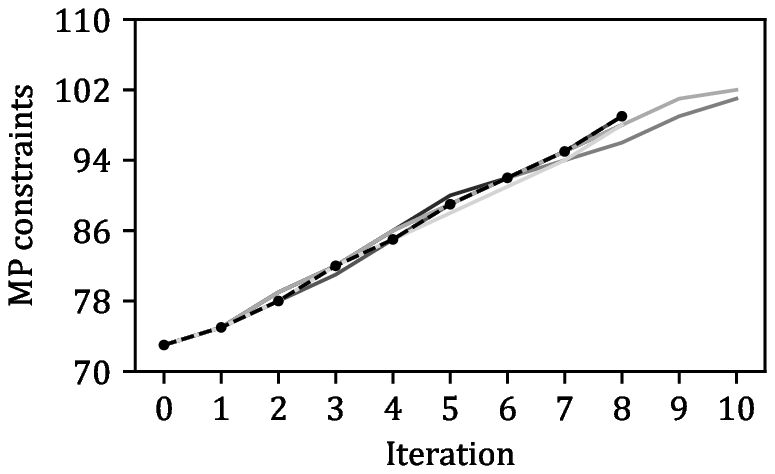}}
  \caption{Number of MP constraints in each iteration of the HQ-MCMS algorithm applied to the 8-bus system for different cut selection strategies and criteria. The trajectories corresponding to each of the 5 runs using a QPU for cut selection are represented by the gray lines. The optimal solution trajectory found using classical optimization is marked with the dashed black line.}
  \label{fig:RMP_size_N8} 
\end{figure*}

\begin{table}[b]
  \centering
  \caption{Largest QUBO Problem Instance Submitted to the QPU for Each Case (8-bus System)}
    \begin{tabular}{lccc}
    \toprule
      & \textbf{Decision variables} & \textbf{Slack variables} & \textbf{Interactions} \\
    \midrule
    \textbf{C7} & 5 & 16 & 99 \\
    \textbf{C8} & 7 & 0 & 21 \\
    \textbf{C9} & 4 & 40 & 186 \\
    \textbf{C10} & 15 & 18 & 176 \\
    \textbf{C11} & 28 & 21 & 336 \\
    \textbf{C12} & 48 & 83 & 763 \\
    \bottomrule
    \end{tabular}%
  \label{tab:N8_problemsize}%
\end{table}%

In Table \ref{tab:results_8bus} it can be seen that all multi-cut strategies result in a reduction in the number of iterations with respect to the straightforward implementation of BD of up to 69.6\%. Since the 8-bus power system problem instance is small, adding all the generated cuts at each iteration to the MP results in the lowest total execution time, despite the increase in the MP solution time. The reason for this is that all cases in which cut selection is applied, additional components are executed, namely the construction of the indicator matrix $\mathbf{M}$ and the solution of the cut selection problem. For this small problem instance, the significant reduction in the DSP solution time (which in turn depends on the number of iterations) is dominant. Nonetheless, applying any of the cut selection strategies and criteria results in a comparable acceleration in comparison with the straightforward implementation of BD when classical resources are used. 

For the 8-bus system, the total cut selection time is by up to 50\% lower using QA for cases C7-C12 in comparison with their classical counterparts C1-C6. However, when QA is used, additional time is required in order to map the problem to the QPU topology, which can significantly influence the performance of the overall performance of the HQC-MCMS algorithm. It can be seen that when Strategy I is used for cut selection, significantly less time is required in order to find a minor embedding in comparison with \mbox{Strategy II.} Also, applying cut selection using $D^{F}$ resulted in faster minor embedding in comparison with $D^{O}$. Nonetheless, for cases C7-C11 the HQC-MCMS algorithm performed better than the straightforward implementation of BD, while C8 also performed better than its classical counterpart C2. It is to be noted that although, on average, the total solution time that is reported in C12 exceeds that of the straightforward implementation of BD, the performance of the minor embedding heuristic is highly variable (standard deviation of \mbox{50.31 s}) and two instances were found to perform better than BD.

A similar analysis is performed for the 30-bus system based on the results that are presented in Table \ref{tab:results_30bus}. First, it can be observed that all the multi-cut solution approaches result in a reduction in the number of iterations with respect to the straightforward BD implementation of up to 83.5\%. Contrary to the 8-bus system, adding all the available cuts in each iteration to the MP results in a proliferation of the MP solution time that is sufficient to render it a less performant option than the application of any of the cut selection strategies when using a classical solver. Both for Strategies I and II the best performance was observed when Criterion II was used by applying cut selection both on feasibility and optimality cuts (C3 and C6) due to the maximum reduction in the size of the MP. However, the corresponding cases using QA (C9 and C12) were the least performant ones due to the excessive time that was required in order to map the corresponding problems to the QPU topology. 

Similarly to the results for the 8-bus power system, except for cases C11 and C12, cut selection was faster when QA was used also for the 30-bus power system by up to 40\%. The worse performance of cut selection in C11 and C12 can be attributed to the higher average number of iterations that were required for convergence to the specified tolerance in comparison with their classical counterparts (C8 and C9). Despite the significant time that is spent on minor-embedding, C7 and C10 are characterized by a lower solution time in comparison with adding all the available cuts in the MP, while C8 was slightly faster than the straightforward BD implementation. \looseness=-1 

The results on the 30-bus test system also reveal a significant difference in the time that is required for minor-embedding the logical problem graphs to the QPU topology when different cut selection strategies and cut selection criteria are applied. Minor embedding time exhibits strong dependence on the complexity of the cut selection problem that is solved using QA, as well as on the size of matrix $\mathbf{M}$ after inspection. Strategy II relies on a more complex problem in comparison with \mbox{Strategy I}, which explains why the total minor embedding is more time consuming in C8 and C9 as well as C11 and C12 in comparison with C7 and C10 respectively. The size of matrix $\mathbf{M}$ is generally larger for optimality cuts in comparison with feasibility cuts. This may be attributed to the different density of feasibility and optimality cuts, which is higher for the latter. For instance, for the worst-performing instance of C9, the lowest average density of feasibility cuts was 7.92\% and the highest 12.50\%, whereas for the optimality cuts the lowest average density was 35.83\% and the highest 55.65\%.

Finally, both for the 8-bus and the 30-bus test systems, HQC-MCMS converged after a significantly lower number of iterations compared to the random cut selection benchmark. This is an indication that despite the fact that QA acts as a heuristic, the cut selection results are non-trivial.

\begin{figure*}[t] 
\centering
  \subfigure[Strategy I, Criterion I (C7)]{\includegraphics[scale=0.7]{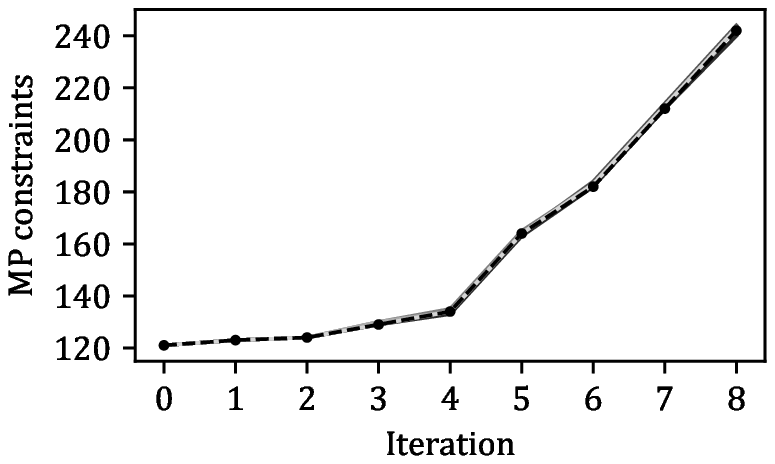}}
  \subfigure[Strategy I, Criterion II, \textit{optSelect}=False (C8)]{\includegraphics[scale=0.7]{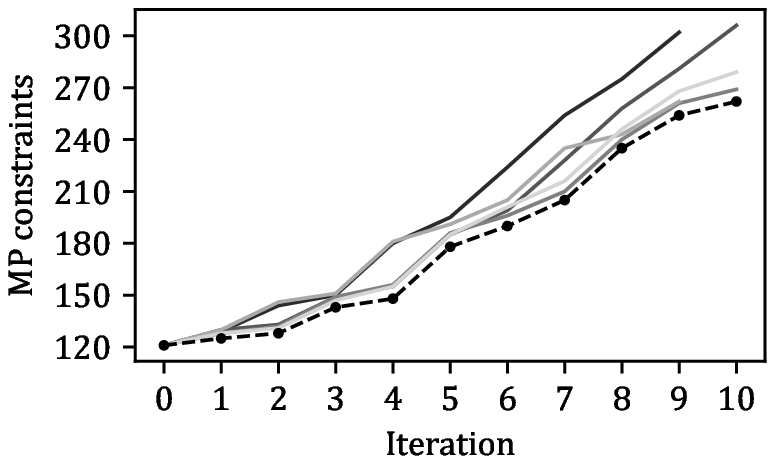}}
  \subfigure[Strategy I, Criterion II, \textit{optSelect}=True (C9)]{\includegraphics[scale=0.7]{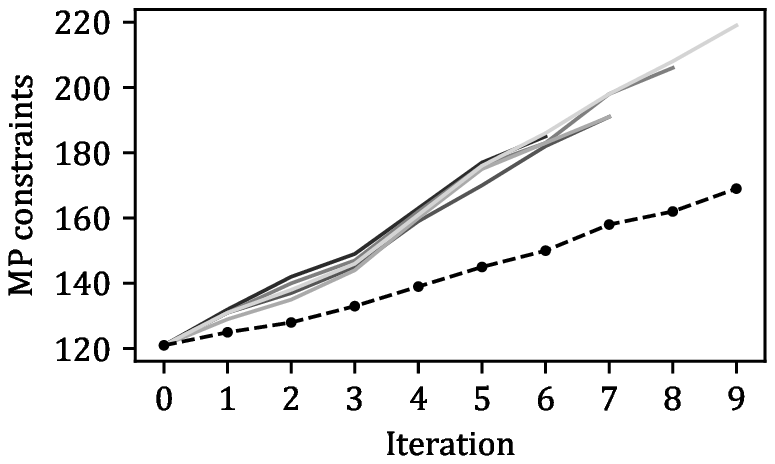}}
  \subfigure[Strategy II,Criterion I (C10)]{\includegraphics[scale=0.7]{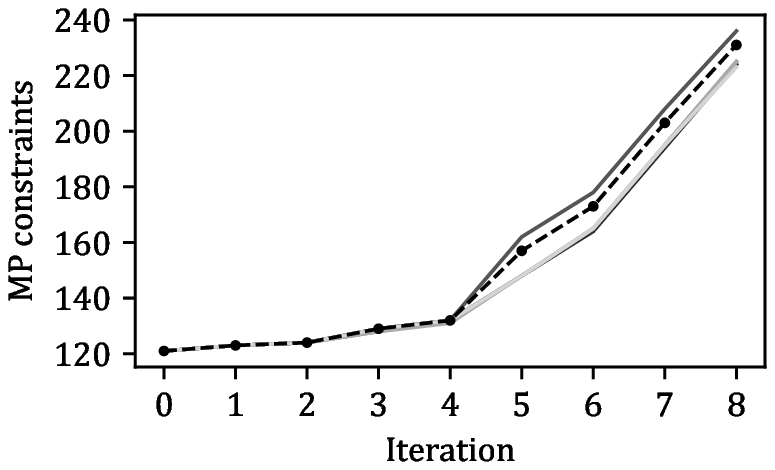}}
  \subfigure[Strategy II, Criterion II, \textit{optSelect}=False (C11)]{\includegraphics[scale=0.7]{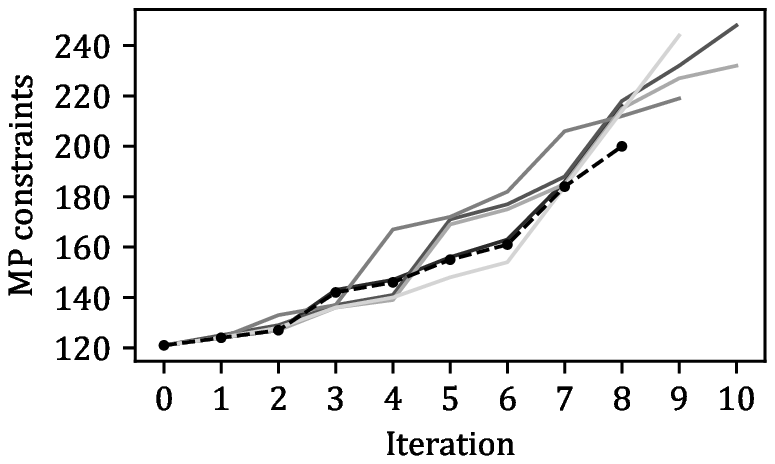}}
  \subfigure[Strategy II, Criterion II, \textit{optSelect}=True (C12)]{\includegraphics[scale=0.7]{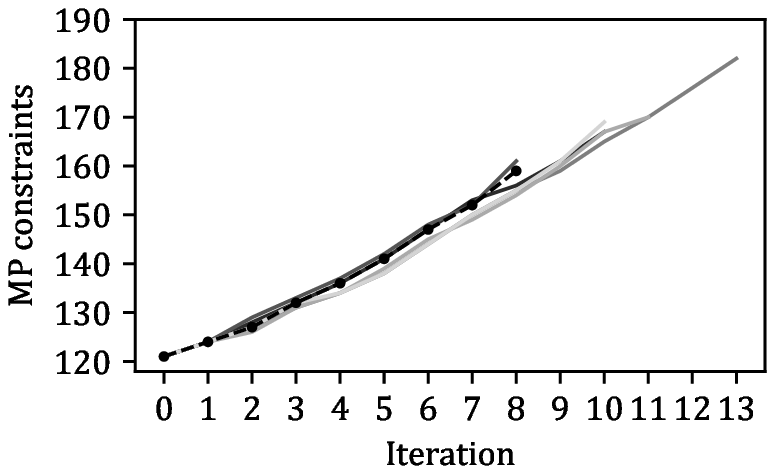}}
  \caption{Number of MP constraints in each iteration of the HQ-MCMS algorithm applied to the 30-bus system for different cut selection strategies and criteria. The trajectories corresponding to each of the 5 runs using a QPU for cut selection are represented by the gray lines. The optimal solution trajectory found using classical optimization is marked with the dashed black line.}
  \label{fig:RMP_size_N30} 
\end{figure*}

\begin{table*}[t]
  \centering
  \caption{QA Solution Details for the 30-bus System}
  \setlength\tabcolsep{3pt}
    \begin{tabular}{lc|c|c|c|c|cc|ccc}
    \toprule
      &   & \multicolumn{3}{c|}{\textbf{Minor embedding}} &   & \multicolumn{2}{c|}{\textbf{Worst case infeasibility}} & \multicolumn{3}{c}{\textbf{Largest problem instance}} \\
    \midrule
      & \makecell{\textbf{\#QPU} \\ \textbf{calls}} & \makecell{\textbf{\#Decisions with} \\ \textbf{broken chains}} & \makecell{\textbf{Max. chain} \\ \textbf{break fraction}} &\makecell{\textbf{Representative} \\ \textbf{chain length}} & \makecell{\textbf{\#Infeas.} \\ \textbf{solutions}}& 
      \makecell{\textbf{\#Infeas.} \\ \textbf{constraints}}& \textbf{\#Constraints} & \makecell{\textbf{Decision} \\ \textbf{variables}} & \makecell{\textbf{Slack}\\\textbf{variables}} & \textbf{Interactions} \\
    \midrule
    \textbf{C7} & 15 & 1 & 1.69 & 5 & - & - & - & 5 & 54 & 324 \\
    \textbf{C8} & 39 & 1 & 1.05 & 7.75 & - & - & - & 17 & 78 & 1559 \\
    \textbf{C9 (feas.)} & 29 & 3 & 1.72 & 7.25 & - & - & - & 18 & 95 & 1937 \\
    \textbf{C9 (opt.)} & 27 & 19 & 5.09 & 34.2 & 2 & 1 & 43 & 30 & 287 & 9569 \\
    \midrule
    \textbf{C10} & 15 & 0 & - & 9.33 & - & - & - & 35 & 52 & 494 \\
    \textbf{C11} & 35 & 9 & 4.83 & 14 & 4 & 1 & 33 & 69 & 99 & 1764 \\
    \textbf{C12 (feas.)} & 45 & 12 & 7.35 & 13.71 & 5 & 2 & 22 & 117 & 149 & 3307 \\
    \textbf{C12 (opt.)} & 34 & 17 & 6.42 & 27.87 & 1 & 1 & 28 & 109 & 165 & 7020 \\
    \bottomrule
    \multicolumn{11}{r}{\#: number}
    \end{tabular}%
  \label{tab:30bus_quality}%
\end{table*}%

\subsubsection{Quality of the solutions obtained using QA}

To evaluate the quality of the solutions that are returned by QA, the increase in the number of constraints of the MP due to the addition of feasibility and optimality cuts across iterations is displayed for the 8-bus and 30-bus test systems in Figs. \ref{fig:RMP_size_N8} and \ref{fig:RMP_size_N30} respectively. \looseness=-1 

For the 8-bus system it can be seen that the increase in the number of MP constraints when cut selection problem is solved using QA corresponds to that of the classical solution for cases C7-C9. This is the reason why no significant differences are observed in the MP solution time in comparison with their classical counterparts, while the same number of iterations are performed. On the contrary, differences are observed when Strategy II is employed in combination with any of the cut selection criteria (C10-C12). Although in all three cases there are instances of the quantum step which result in an optimal cut selection trajectory, the average number of iterations is increased due to the sub-optimal trajectories that are generated in some of the runs. It should be noted that the size of the problems that are submitted to the QPU is relatively small, with the largest instance solved using QA in each case reported in Table \ref{tab:N8_problemsize}. For this reason, the lowest-energy solutions tend to satisfy the constraints of the cut selection strategies. It is also worth mentioning that for all the problems related to this test system that were submitted to the QPU, only a single decision was made based on a sample with broken logical chains (chain break fraction of 0.76\%). The comparable performance of the classical and quantum resources shows that HQC-MCMS is a viable decomposition-based HQC algorithm for small-scale optimization problems. 

For the 30-bus test system, the results portrayed in Fig. \ref{fig:RMP_size_N30} are indicative of the performance of QA as a heuristic for cut selection when HQC-MCMS is applied to larger-scale optimization problems. For Strategy I, although in C7 the application of QA resulted in the selection of a comparable number of cuts in each iteration with its classical counterpart, using \mbox{Criterion II} results in trajectories that significantly increase the size of the MP. This is reflected on the increased MP solution time but also on the reduced average number of iterations required for convergence in C8 and C9. For Strategy II, the opposite behavior is observed. With the exception of C10 in which the results are consistent with the optimal results obtained by classical optimization, C11 and C12 are characterized by a higher number of iterations in comparison with their classical counterparts, which is an indication that a larger number of sub-optimal cut choices are made by QA. Nonetheless, despite the cut selection trajectories generated by QA in C9 and C12 departing significantly from the optimal, the cut selection remains effective in terms of MP size management for Criterion II applied both to feasibility and optimality cuts. \looseness=-1 

To provide further insight into the quality of the QA solutions for the larger 30-bus system, additional details for different cut selection strategies and criteria are presented in Table \ref{tab:30bus_quality} for all the executions of the HQC-MCMS using the QPU. The first observation is that the deterioration of the solution quality observed in Fig. \ref{fig:RMP_size_N30} is related to the size of the QUBO instances that are assigned to the QPU. Furthermore, the size of the problem impacts the qualitative characteristics of minor-embedding. Larger problem instances tend to result in more decisions being made based on samples with broken logical chains and higher maximum chain break fractions. The minor embedding of larger QUBO problems also tends to be characterized by longer qubit chains. A representative chain length is also presented, which is significantly higher when Criterion II is used, in particular when cut selection is applied to optimality cuts. For each minor embedding that is calculated if cut selection is triggered, the maximum qubit chain length is recorded and the results are averaged for each of the five runs. Their maximum across the five runs is considered as a representative chain length. Notably, QA managed to discover valid solutions in the majority of the cases, whereas in the few cases where a cut selection was made based on an infeasible solution, only a few of the problem constraints were not satisfied. Summarizing \mbox{Table \ref{tab:30bus_quality},} \mbox{Strategy I} is associated with more favorable solution characteristics than \mbox{Strategy II.} The same can be said about Criterion I in comparison with Criterion II, especially when cut selection is applied to optimality cuts. \looseness=-1

\subsubsection{Practical limitations}\label{limitations}

Based on the computational experience with the UC problem, three limitations of the proposed HQC algorithm can be identified. First, although Criterion II appears to be computationally advantageous in comparison with Criterion I when classical computing resources are used, the dependence of the size of matrix $\mathbf{M}$ on the number of complicating variables imposes a limit on the size of the problem that can be embedded on the QPU. In addition to that, significant time is consumed on minor-embedding, especially when cut selection concerns optimality cuts. On the contrary, the size of matrix $\mathbf{M}$ under Criterion I depends solely on the number of MP solutions which is a user-selected parameter. Since this parameter is independent of the size of the optimization problem, Criterion I may find wider applicability to large-scale optimization problems when using a QPU to perform cut selection in comparison with \mbox{Criterion II}, given the limitations of NISQ hardware. 

Another shortcoming of HQC-MCMS is that minor embedding has to be repeated in each iteration where the QPU is called because a different matrix $\mathbf{M}$ is available. Although performing cut selection using quantum resources may be a computationally efficient procedure itself, the impact of the time that is required in order to map the problem can adversely impact the overall performance of the algorithm. To overcome this limitation, either more efficient minor-embedding heuristics or a systematic way to exploit previously generated minor embeddings need to be developed. 

The third challenge is related to finding suitable values for the hyperparameters (chain strength, annealing schedule) and weights for the quadratic penalty functions in order to guarantee that the lowest-energy solution that is found corresponds to a valid solution with high probability and reduce logical chain breakage.

\section{Conclusion}\label{lab:conclusion}
In this paper, a hybrid quantum-classical (HQC) multi-cut Benders decomposition (BD) strategy that exploits multiple feasible solutions of the master problem (MP) in order to generate multiple feasibility and optimality cuts was presented. Adding multiple cuts to the MP improves the convergence rate of BD. However, the increase in the size of the MP may adversely impact solution time. In order to manage the size of the MP and exploit the availability of multiple cuts, a cut selection procedure that can be assigned to a quantum computer was developed. Two different criteria and two different cut selection strategies based on pure binary problems that can be solved using quantum computing were studied. The HQC algorithm was applied to the Unit Commitment problem and computational experiments were conducted using the \mbox{D-Wave} \mbox{Advantage 4.1} quantum annealer. Results on two test power systems showed that although it is viable for quantum resources to be used as an alternative to classical resources for cut selection for small-scale problems, current hardware limitations must be overcome and the efficiency of minor-embedding techniques should be improved before effectively applying the proposed approach to large-scale problem instances. Future research will focus on further improving the proposed HQC algorithm according to the limitations that were identified in \mbox{Section \ref{limitations}} and applying it on different use cases. \looseness=-1

\bibliographystyle{IEEEtran}
\bibliography{bibliography/bibliography}




\end{document}